\documentclass[
 floatfix,
 reprint,
 amsmath,amssymb,
 aps,
]{revtex4-2}
\usepackage[utf8]{inputenc}
\usepackage{amsmath}
\usepackage[table,xcdraw]{xcolor}
\usepackage{graphicx}
\usepackage{hyperref}
\usepackage{import}
\usepackage{svg}
\usepackage{dcolumn}
\usepackage{physics}
\usepackage{bm}
\usepackage{tikz}
\usepackage{amsthm}
\usepackage{tabularx,booktabs}
\usepackage[toc,page]{appendix}
\usepackage[export]{adjustbox}
\usepackage{siunitx}
\usepackage{upgreek}

\usetikzlibrary{tikzmark}
\usetikzlibrary{braids}
\newcolumntype{Y}{>{\centering\arraybackslash}X}

\newtheorem{theorem}{Theorem}

\newcommand{\bn}{\mathbf{n}}
\newcommand{\blambda}{\bm{\lambda}}
\newcommand{\bchi}{\bm{\chi}}
\newcommand{\btau}{\bm{\tau}}
\newcommand{\jmt}{~\textrm{J}/\textrm{m}^3}
\newcommand{\pN}{~\textrm{pN}}
\newcommand{\um}{~\upmu \textrm{m}}
\newcommand{\hide}[1]{}

\begin{document}

\title{Topological Rigidity and Non-Abelian
defect junctions in chiral nematic systems with effective biaxial symmetry }

\author{Jin-Sheng Wu}
\thanks{These two authors contributed equally}
\affiliation{Department of Physics and Chemical Physics Program, University of Colorado, Boulder}

\author{Roberto Abril Valenzuela}
\thanks{These two authors contributed equally}
\affiliation{Department of Physics, University of California, Santa Barbara}

\author{Mark J. Bowick}
\email{bowick@kitp.ucsb.edu}
\affiliation{Kavli Institute for Theoretical Physics, University of California Santa Barbara
\\
French American Center for Theoretical Science, CNRS, KITP, Santa Barbara\hide{, California 93106-4030, USA}}

\author{Ivan I. Smalyukh}
\email{ivan.smalyukh@colorado.edu}
\affiliation{Department of Physics and Chemical Physics Program, University of Colorado, Boulder\\International Institute for Sustainability with Knotted Chiral Meta Matter\\ Department of Electrical, Computer, and Energy Engineering,
Materials Science and Engineering Program and Soft Materials Research Center, University of Colorado, Boulder\\ Renewable and Sustainable Energy Institute, National Renewable Energy
Laboratory and University of Colorado, Boulder}


\begin{abstract}
    We study topologically stable defect structures in systems where the defect line classification in three dimensions and associated algebra of interactions (the fundamental group) are governed by the non-Abelian 8-element group, the quaternions $Q_8$. The non-Abelian character of the defect algebra leads to a topological rigidity of bound defect pairs, and trivalent junctions which are the building blocks of multi-junction trivalent networks. We realize such structures laboratory chiral nematics and analyze their behavior analytically, along with numerical  modeling
\end{abstract}

\maketitle

\section{Introduction}

Topological defects of various intrinsic dimensions, from superfluid filaments to the atmospheric polar vortex, play important roles in a wide variety of physical systems~\cite{mermin_topological_1979,kleman1989defects, de1992defect, ovidko1987topological, nelson2002defects}. Dislocation motion and entanglement are key determinants of the mechanical response of solids to applied stresses and superconducting vortices, pinned or unpinned,  are a critical source of dissipation in superconductors. Topological defects are also essential singularities in the sense that, provided topology allows them, they inevitably form in finite-rate continuous phase transitions due to critical slowing down and, being the slowest degrees of freedom, they control the rate of such transitions \cite{goldenfeld1995dynamics, Kibble1976cosmic, kibble1995phase}. They are distinctive fingerprints of phase transitions involving spontaneously broken symmetries in that the specific topological classes they form often distinguish different symmetry-breaking patterns and the associated order-parameter spaces. 

In most cases, and especially in three dimensions, the pure interactions of the minimal-energy defects are governed by an Abelian (commutative) symmetry structure such as a discrete $\mathbb{Z}_n$ or the group of integers $\mathbb{Z}$~\cite{mermin_topological_1979, nelson2002defects}. These minimal-energy defects only interact at a distance, or indirectly through their coupling to other material fields~\cite{kleman1989defects}. Defects then have only energetic barriers to crossing and do not form stable defect junctions or networks in the absence of pinning centers.
This is the case for superfluid filaments, superconducting flux lines and uniaxial three-dimensional liquid crystals (nematics), as well as elastic disclinations and dislocations. 

Both biaxial liquid crystals and liquid crystals with intrinsic twist (chiral nematics) have a rich class of topological line defects that are, instead, governed by a non-Abelian (non-commutative) algebraic structure, leading to junctions where three defects meet, the physical entanglement of defects and history-dependent interactions. Here we show that all these features can be experimentally realized in a laboratory chiral nematic and agree well with theoretical predictions and the results of numerical simulations.  

We experimentally visualize and demonstrate that a conventional chiral nematic displays all the key topological features expected from the non-Abelian algebra of its associated defects, including two distinct types of trivalent junction, topological entanglement of defect pairs and stable parallel defect lines. 

After introducing the key background material in Secs. II-V~\cite{mermin_topological_1979,kleman1989defects, de1992defect, ovidko1987topological, nelson2002defects}, and following the methods of Sec.VI, we exhibit the two basic types of trivalent defect junctions: three distinct defects meeting at a point, or two defects from the same class producing or annihilating a non-trivial radial defect - the so-called $-1$ defect. The $-1$ defect is also necessarily generated when two braided defects of the same class, fixed at their ends, are pulled across each other, thus providing a topological obstruction to disentanglement, or said differently, generating topological rigidity for a braided pair. Such a pair is then a fundamental excitation in the system, analogous to a topologically protected and therefore stable bound state~\cite{mermin_topological_1979,kleman1989defects, de1992defect, ovidko1987topological, nelson2002defects}. 

We then show that trivalent junctions can be joined to create an extended flexible and stable trivalent network, 
a filamentous and fluid fishnet, paving the way for creating a wide variety of larger-scale networks.
By experimentally patterning defects of a fixed class at the surface of the experimental cell we show that disclination lines can form handles connecting the surface to itself and bridges connecting one surface to another, all controlled by the relative sign of the topological class of the surface defects.

Finally, we analytically treat the case of two parallel disclination lines where intrinsic defect repulsion is balanced by the need to minimize the energy in the twist of the chiral nematic, again leading to a stable bound state observed in our experiments.

From a broader perspective, our results apply to all three-dimensional, rotationally symmetric liquid crystal systems with an isotropy given by the dihedral group ($D_2$). This occurs whenever there is a well-defined orthorhombic frame, either local or global. The global case arises in biaxial nematic liquid crystals formed from brick-like building blocks \cite{freiser1970ordered, liu2016, mundoor2021, mundoor2018,smalyukh2020Review, tschierske2010biaxial, luckhurst2015biaxial}, while the emergent local counterpart case occurs in chiral nematics (cholesterics) and in hybrid molecular/colloidal systems, where the symmetry breaking in molecular and colloidal order relies crucially on the presence of a helical axis \cite{liu2016, priest1974biaxial,wu2023unavoidable, luckhurst2015biaxial}. For clarity of notation, we will call the entire class D2CLC (order 2 dihedral chiral liquid crystals).

D2CLC systems can be realized experimentally by exploiting the biaxiality of a chiral liquid crystal (CLC), as implemented here, or in hybrid molecular-colloidal systems (colloidal elements in a liquid crystal environment~\cite{wu2023unavoidable, liu2016, mundoor2021, mundoor2018,smalyukh2020Review, tschierske2010biaxial}. The associated topological defects are at least partially classified by the homotopy classes of the order parameter manifold, which specifies the space of inequivalent ground states. For D2CLC the first homotopy group of the space of ground states, known as the fundamental group, is the non-Abelian 8-element group, the quaternions $Q_8$ ~\cite{toulouse_pour_1977}, first introduced by Hamilton in 1843 in an attempt to find a three-dimensional version of the complex plane~\cite{mermin_topological_1979,kleman1989defects}. The quaternionic fundamental group leads to five classes (conjugacy classes) of topologically stable line defects in three dimensions, one trivial class (the identity), and four non-trivial classes. The quaternionic algebra has several important consequences. Braids of two defects from distinct classes are topologically entangled ~\cite{poenaru_crossing_1977,mermin_topological_1979, alexander2022entanglements} and can only be disentangled by creating a third distinct bridging defect. Perhaps more importantly there are stable trivalent junctions where three defects meet, allowing the spontaneous formation of extended line-defect networks consisting of nodes (junctions) of coordination number (degree) three. \cite{poenaru_crossing_1977}.


While quantum entanglement and braiding of anyons in hard condensed matter systems with topological order in two spatial dimensions is very rich~\cite{google2023non, iqbal2024non, nayak2008non}, we explore here a different mechanism for physical entanglement and braiding in three spatial dimensions, based solely on the non-trivial topology of the ground state manifold.

\section{Biaxial models of chiral nematics} \label{sec:BiaxialElasticity}

\begin{figure*}
        \centering 
        \includegraphics[width=\textwidth]{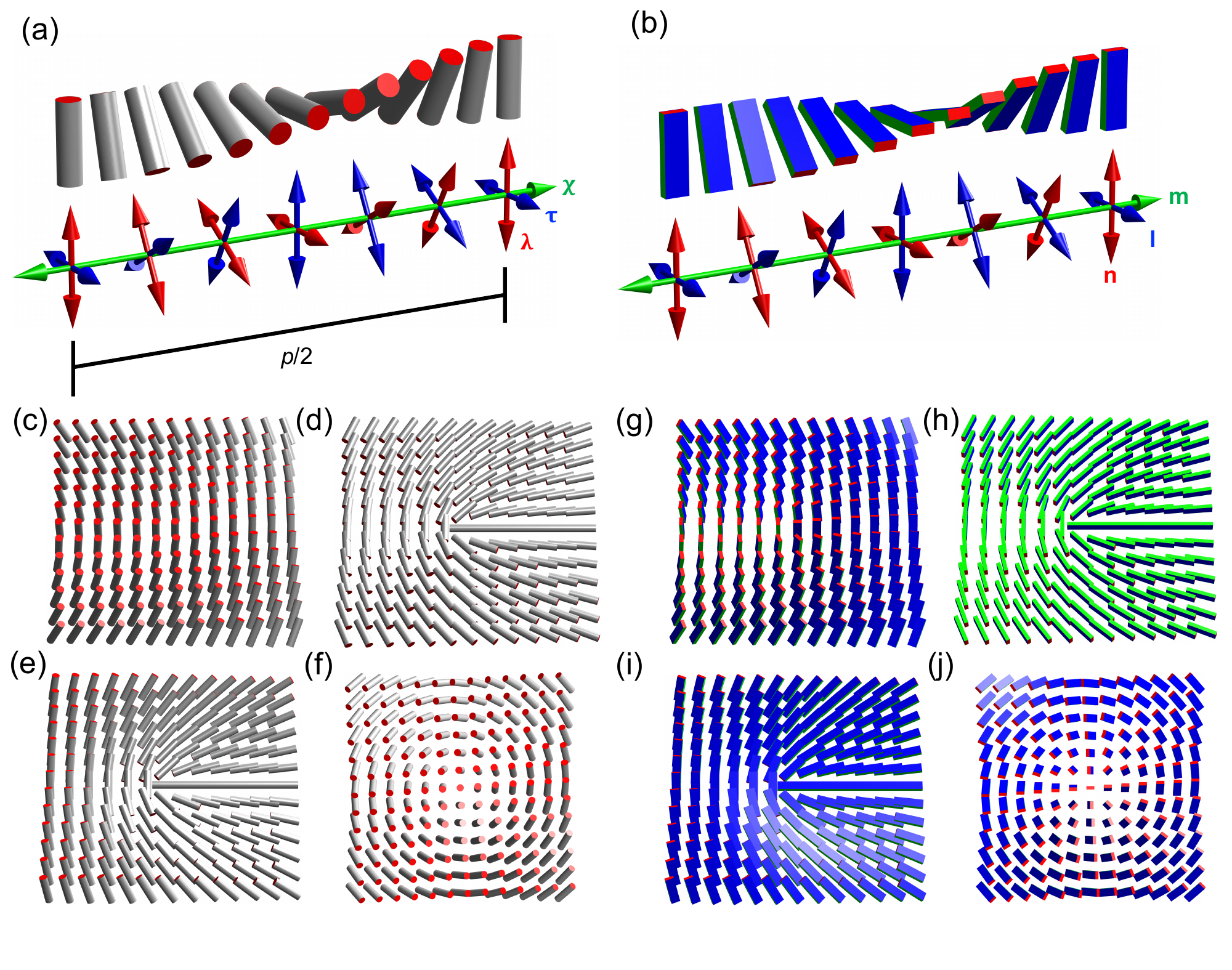}
        \caption{(a-b) Field definition in the (a) chiral nematic setting and (b) respective biaxial fields under the mapping established in section II and Appendix C. The cholesteric pitch $p$ is defined as the distance for a nematic constituent director $\blambda$ to rotate by $2\pi$ along helical axis $\bchi$ as shown in (a). The corresponding fields describing biaxiality, $(\bn, \mathbf{m}, \mathbf{l})$, define the biaxial triad parallel to brick edges in (b). (c-f) Cross-sections of line defect textures in chiral nematics belonging to $C_\lambda$, $C_\chi$, $C_\tau$, and $-1$ defect classes, respectively. (g-j) Corresponding biaxial ``brick" textures showing defects are nonsingular in $\bn$, $\mathbf{m}$, $\mathbf{l}$, and $\bn$ respectively, while being singular in the 2 other fields orthogonal to them. Here we only show one of the three representatives of the $-1$ defect class. } 
     \label{fig:brickmap}
\end{figure*}

Chiral liquid crystals are nematic systems with broken mirror symmetry, which manifests itself as helicoidal twist configurations in the {\it apolar} nematic director, $\blambda(\mathbf{r})=-\blambda(\mathbf{r})$. Locally, the twist defines a helical twist axis $\bchi=-\bchi$, orthogonal to $\blambda$, providing a lengthscale, $p/2$, over which $\blambda$ rotates by $\pi$, forming one cholesteric quasi-layer (Fig.~\ref{fig:brickmap}(a)). The local set of fields $(\blambda,\bchi,\btau)$, derived from the twist, defines the molecular field, helical axis  field, and a third orthogonal direction ($\btau=\blambda\times\bchi$), respectively (see Fig.~\ref{fig:brickmap}(a)). This establishes a local triad of orthogonal directors and leads to a connection between CLCs and orthorhombic biaxial systems described by three mutually orthogonal fields $(\mathbf{n},\mathbf{m},\mathbf{l})$. Conceptually, the non-degeneracy of $(\blambda,\bchi,\btau)$ (and most importantly the latter two) already breaks uniaxial symmetry because of the intrinsic biaxial features of twist alignment. Physically, the orientational distributions of molecular orientations relative to $\bchi$ and $\btau$ are different ~\cite{wu2023unavoidable, priest_biaxial_1974}. 
To see the actual mapping between chiral and biaxial nematics, however, one needs to examine the free energy models of CLCs from the perspective of biaxial nematics. Firstly, it is straightforward that the molecular orientation is identical to the nematic director $\blambda\equiv\bn$. The remaining  mapping from the chiral nematic frame $(\blambda,\bchi,\btau)$ to the biaxial frame $(\mathbf{n},\mathbf{m},\mathbf{l})$ is  revealed by the tensorial model of CLC free energies detailed below.

In the tensorial model, the orientational order in a CLC is encoded in the traceless, symmetric tensor order parameter $Q_{ij}$, given in terms of the directors $\mathbf{n}$ and $\mathbf{m}$ by:
    \begin{equation}
        Q_{ij} = S\left(n_i n_j-\frac{\delta_{ij}}{3}\right) + T\left(m_i m_j-\frac{\delta_{ij}}{3}\right) ,        \label{Qtensor}
    \end{equation}
where $S$ and $T$ are the uniaxial and biaxial orientational order parameters, respectively, and $\delta_{ij}$ is the Kronecker-delta.
One may write the strain-free elastic free energy density as a sum of derivatives of $Q_{ij}$~\cite{vissenberg1997generalized, govers_elastic_1984, priezjev_coarsening_2002}:
    \begin{align}
        f_{\mathrm{elastic}} &= \gamma_1 \partial_k Q_{ij}\partial_k Q_{ij} + \gamma_2 \partial_j Q_{ij} \partial_{k}Q_{ik} \nonumber\\
        &+ \gamma_6Q_{ij}\partial_i Q_{kl}\partial_j Q_{kl} \  .
        \label{Qenergy}
    \end{align}
Another second-order term $\gamma_3\partial_k Q_{ij}\partial_j Q_{ik}$ is sometimes considered in the tensorial model. In our analysis of the bulk elastic energies, however, $\gamma_3$ differs from $\gamma_2$ only by a surface integral and so this term is dropped~\cite{vissenberg1997generalized}.
A 3rd-order term is also necessary to incorporate anisotropy of the elastic contributions ~\cite{lubensky1970molecular,longa_extension_1987}. Eq.~\ref{Qenergy}
is stable for positive order parameters $S>T>0$.
Chirality may be introduced, to lowest order, by adding a term of the form 
    \begin{equation}
        f_{\mathrm{chiral}} = \gamma_4 \epsilon_{ijk} Q_{il}\partial_{j} Q_{kl} 
        \label{ChiralTerm} \ ,
    \end{equation}
where $\epsilon_{ijk}$ is the Levi-Civita tensor~\cite{longa_extension_1987}. 


The diagonalization of the $\mathbf{Q}$ tensor sets the orthonormal triad $(\mathbf{n},\mathbf{m},\mathbf{l})$, as in Eq.~\ref{Qtensor} (see Methods). 
On the other hand, by identifying the molecular axes and the principal director $\blambda\equiv\bn$, one can also reconstruct the chiral directors $(\blambda,\bchi,\btau)$, based on methods connected to the twist in $\bn$~\cite{beller_geometry_2014,wu2023unavoidable,wu2022hopfions, ZhangPRL}. 
For a variety of CLC structures derived from minimizing the free energy Eq.~\ref{Qenergy} and Eq.~\ref{ChiralTerm}, we find that the helical axis $\bchi$, along which $\blambda$ rotates, is consistently aligned with the direction of the biaxial director $\bchi\equiv\mathbf{m}$ (Fig.~\ref{fig:brickmap}). 
In our CLC systems then, where twist is the dominant elastic deformation, the biaxiality comes mainly from minimizing the chiral term Eq.~\ref{ChiralTerm}.
Lastly, the transverse molecular axis is the cross product of the other two directors and represents the third axis $\btau\equiv\mathbf{l}$, and the mapping of chiral nematic directors to orthorhombic biaxial directors is valid for CLC systems with a consistent twisting handedness.

Fig.~\ref{fig:brickmap}(a-b) shows the comparison of the chiral and biaxial field definitions, indicating the direct mapping of the molecular and helical axes. 
In a variety of structures, such as Fig.~\ref{fig:brickmap}(c-j), one also finds a degree of biaxiality $T \propto q^2$, with chirality $q=2\pi/p$ the principal eigenvalue of the handedness tensor~\cite{priest_biaxial_1974,wu2023unavoidable}, suggesting an interchangeable interpretation of biaxiality and chirality in CLCs based on their symmetries. In all this leads to a biaxial structure  coupled to that of the helical configuration of the CLC \cite{priest_biaxial_1974,wu2023unavoidable}.
In fact, any deformation (not just twist) in the material director field $\blambda$ causes weak biaxiality and, thus, the biaxial tensor field is always well defined unless the helical structure is fully unwound and the material director field is spatially uniform ~\cite{wu2023unavoidable}. The biaxial description of CLCs based on our mapping is thus more robust when applied to the analysis of defects (Methods)~\cite{wu2023unavoidable}. 

The mapping of the chiral and biaxial director fields allows us to use the biaxial  free energy, written in terms of derivatives of the directors, $(\mathbf{n},\mathbf{m},\mathbf{l})$, to describe a CLC. It has been shown that one can express this as a sum of 12 linearly-independent bulk elastic terms~\cite{govers_elastic_1984}: 
\begin{align}
    &f_{\mathrm{elastic}}^{\mathrm{FO}} = \frac{K_1}{2} (\nabla\cdot \mathbf{n})^2 +\frac{K_2}{2} (\mathbf{n}\cdot\nabla\times \mathbf{n})^2 + \frac{K_3}{2}(\mathbf{n}\times\nabla\times\mathbf{n})^2 \nonumber\\
    &+\frac{K_4}{2} (\nabla\cdot \mathbf{m})^2 +\frac{K_5}{2} (\mathbf{m}\cdot\nabla\times \mathbf{m})^2 + \frac{K_6}{2}(\mathbf{m}\times\nabla\times\mathbf{m})^2 \nonumber \\\
    &+ \frac{K_7}{2}\left[\mathbf{n}\cdot(\mathbf{m}\times\nabla\times\mathbf{m})\right]^2 + \frac{K_8}{2}\left[\mathbf{m}\cdot(\mathbf{n}\times\nabla\times\mathbf{n})\right]^2  \nonumber\\
    &+ \frac{K_9}{2}\left[\mathbf{m}\cdot\nabla\times\mathbf{l}\right]^2 + \frac{K_{(10)}}{2}\left[\mathbf{n}\cdot\nabla\times\mathbf{l}\right]^2 \nonumber \\
    &+ \frac{K_{(11)}}{2}(\nabla\times\mathbf{l})^2 + \frac{K_{(12)}}{2} (\nabla\cdot\mathbf{l})^2 \ ,
    \label{FrankOseenE}
\end{align}
with the first three terms resembling the uniaxial Frank-Oseen model.
In this way the chirality contribution to orientational elasticity is expressed as
    \begin{align}
        f_{\mathrm{chiral}}^{\mathrm{FO}} &= 
        K_{(13)} (\mathbf{n}\cdot\nabla\times \mathbf{n}) + 
        K_{(14)} (\mathbf{m}\cdot\nabla\times \mathbf{m}) \nonumber \\
        &+ K_{(15)} (\mathbf{n}\times\mathbf{m})\cdot(\mathbf{m}\cdot\nabla)\mathbf{n} \nonumber \\
        &+ K_{(16)} (\mathbf{m}\times\mathbf{n})\cdot(\mathbf{n}\cdot\nabla)\mathbf{m} \nonumber \\
        &+ K_{(17)}(\mathbf{n}\times\mathbf{m})\cdot(\mathbf{m}\times\nabla\times\mathbf{n}-\mathbf{n}\times\nabla\times\mathbf{m}).
        \label{FOchiral}
    \end{align}

Relating the two elastic models by expanding Eq.~\ref{Qenergy} and comparing to Eq.~\ref{FrankOseenE}, one can write the elastic constants $K_i$ in terms of the constants $\gamma_i$. This yields a biaxial description of the elastic properties of chiral nematics (see Appendix C). Clearly the elastic constants $K_i, i=4-12$ scale as $\mathcal{O}(T)$, so that vanishing biaxiality reduces to the conventional uniaxial Frank-Oseen model with only the first three elastic terms. 
On the other hand, the chirality-amplified biaxiality~\cite{priest_biaxial_1974,harris1997microscopic,kroin1989chirality,longa1994biaxiality} gives rise to elastic contributions beyond bend, twist or splay in the molecular director field.
We estimate the values of elastic constants $K_i, i=4-12$ for chiral nematics with the assumption $T\ll S$ (see Appendix C), representing the elastic moduli of orthorhombic biaxial nematics that originate purely from chirality, in addition to the intrinsic chiral deformations within such models. This mapping of the elasticities of CLCs to  biaxial nematics  strengthens the fundamental connection of the two systems, later allowing us to analytically model experimental observations of bound states while using parameters measured in experiments. 





\section{Homotopy analysis of biaxial nematics}

After the possibility of a biaxial phase of liquid crystals was first introduced and analyzed  by Freiser~\cite{freiser1970ordered}, the topological character of the ground state manifold (or order parameter space) associated with symmetry breaking from an isotropic phase to a biaxial phase was analyzed by Toulouse~\cite{toulouse_pour_1977}.  The order parameter space is $SO(3)/D_2$, the space of all $3$-dimensional rotations that obey the symmetry of the rectangle, and the resultant line defects following from its fundamental group are classified by the group of quaternions $Q_8$: $\pi_1[SO(3)/D_2]=Q_8=\{\pm1,\pm i,\pm j,\pm k\}$ ~\cite{nozaki2024homotopy,nakanishi_topological_1988, alexander2022entanglements, annala2022topologically}. Unknotted ring defects were also classified in \cite{nakanishi_topological_1988}. There has been much work in this area in the last 50 years (for a review see \cite{ alexander2012colloquium}). 

In contrast to uniaxial nematics, whose fundamental group algebra is Abelian (the integers in 2-dimensions or $\mathbb{Z}_2$ in three dimensions), the  quaternions are  non-Abelian. In fact, the quaternion group is the smallest non-Abelian group that arises by modding out a discrete group from the rotation group~\cite{mermin_topological_1979}. The multiplication table for the group elements of $Q_8$ is given in Table \ref{tab:group_table}. We proceed with the biaxial analogy of CLCs and relabel the quaternion elements using the fields $\blambda,\bchi,\btau$:  $(i,j,k)\rightarrow(\lambda,\chi,\tau)$~\cite{kleman2003soft}. 

\begin{table}
\begin{tabularx}{\columnwidth}{c| *{8}{Y}}
    $\times$ & $1$                          & $-1$                         & $\lambda$                          & $-\lambda$                         & $\tau$                          & $-\tau$                         & $\chi$                          & $-\chi$                         \\ \hline
$1$  & \cellcolor{blue!25}$1$  & \cellcolor{red!25}$-1$ & \cellcolor{cyan!25}$\lambda$  & \cellcolor{cyan!25}$-\lambda$ & \cellcolor{yellow!25}$\tau$  & \cellcolor{yellow!25}$-\tau$ & \cellcolor{green!25}$\chi$  & \cellcolor{green!25}$-\chi$ \\ \hline
$-1$ & \cellcolor{red!25}$-1$ & \cellcolor{blue!25}$1$  & \cellcolor{cyan!25}$-\lambda$ & \cellcolor{cyan!25}$\lambda$  & \cellcolor{yellow!25}$-\tau$ & \cellcolor{yellow!25}$\tau$  & \cellcolor{green!25}$-\chi$ & \cellcolor{green!25}$\chi$  \\ \hline
$\lambda$  & \cellcolor{cyan!25}$\lambda$  & \cellcolor{cyan!25}$-\lambda$ & \cellcolor{red!25}$-1$ & \cellcolor{blue!25}$1$  & \cellcolor{green!25}$\chi$  & \cellcolor{green!25}$-\chi$ & \cellcolor{yellow!25}$-\tau$ & \cellcolor{yellow!25}$\tau$  \\ \hline
$-\lambda$ & \cellcolor{cyan!25}$-\lambda$ & \cellcolor{cyan!25}$\lambda$  & \cellcolor{blue!25}$1$  & \cellcolor{red!25}$-1$ & \cellcolor{green!25}$-\chi$ & \cellcolor{green!25}$\chi$  & \cellcolor{yellow!25}$\tau$  & \cellcolor{yellow!25}$-\tau$ \\ \hline
$\tau$  & \cellcolor{yellow!25}$\tau$  & \cellcolor{yellow!25}$-\tau$ & \cellcolor{green!25}$-\chi$ & \cellcolor{green!25}$\chi$  & \cellcolor{red!25}$-1$ & \cellcolor{blue!25}$1$  & \cellcolor{cyan!25}$\lambda$  & \cellcolor{cyan!25}$-\lambda$ \\ \hline
$-\tau$ & \cellcolor{yellow!25}$-\tau$ & \cellcolor{yellow!25}$\tau$  & \cellcolor{green!25}$\chi$  & \cellcolor{green!25}$-\chi$ & \cellcolor{blue!25}$1$  & \cellcolor{red!25}$-1$ & \cellcolor{cyan!25}$-\lambda$ & \cellcolor{cyan!25}$\lambda$  \\ \hline
$\chi$  & \cellcolor{green!25}$\chi$  & \cellcolor{green!25}$-\chi$ & \cellcolor{yellow!25}$\tau$  & \cellcolor{yellow!25}$-\tau$ & \cellcolor{cyan!25}$-\lambda$ & \cellcolor{cyan!25}$\lambda$  & \cellcolor{red!25}$-1$ & \cellcolor{blue!25}$1$  \\ \hline
$-\chi$ & \cellcolor{green!25}$-\chi$ & \cellcolor{green!25}$\chi$  & \cellcolor{yellow!25}$-\tau$ & \cellcolor{yellow!25}$\tau$  & \cellcolor{cyan!25}$\lambda$  & \cellcolor{cyan!25}$-\lambda$ & \cellcolor{blue!25}$1$  & \cellcolor{red!25}$-1$ \\ \hline
\end{tabularx}
\caption{Multiplication table of the quaternion group elements $Q_8$ using the chiral nematic fields $\lambda, \tau, \chi$ as relabelings of the elements $(i,j,k)$ . Conjugacy class assignments are highlighted by different colors.}
\label{tab:group_table}
\end{table}

\begin{table}
    \centering
    \begin{tabularx}{\columnwidth}{ p{5.1mm}|p{5.1mm}p{5.1mm}ccc }
        
         & $+1$  & $-1$  & $C_\lambda$       & $C_\tau$       & $C_\chi$       \\ \hline
        $+1$  & $+1$  & $-1$  & $C_\lambda$       & $C_\tau$       & $C_\chi$       \\ \hline
        $-1$  & $+1$  & $+1$  & $C_\lambda$       & $C_\tau$       & $C_\chi$       \\ \hline
        $C_\lambda$ & $C_\lambda$ & $C_\lambda$ & $2(+1) \oplus 2(-1)$ & $2C_\chi$       & $2C_\tau$       \\ \hline
        $C_\tau$ & $C_\tau$ & $C_\tau$ & $2C_\chi$       & $2(+1) \oplus 2(-1)$ & $2C_\lambda$       \\ \hline
        $C_\chi$ & $C_\chi$ & $C_\chi$ & $2C_\tau$       & $2C_\lambda$       & $2(+1) \oplus 2(-1)$ \\ 
        \hline
    \end{tabularx}
    \caption{Multiplication table for the conjugacy classes of $Q_8$. Note that combining two defects belonging to the same class produces an ambiguous result, $2(+1) \oplus 2(-1)$, where the 2 indicates the degeneracy of the result (see Table~\ref{tab:group_table}), in which case the actual result is determined by the path taken to combine the two initial defects.}
    \label{tab:conj_table}
\end{table}

\begin{figure}
        \centering 
        \includegraphics[width=\columnwidth]{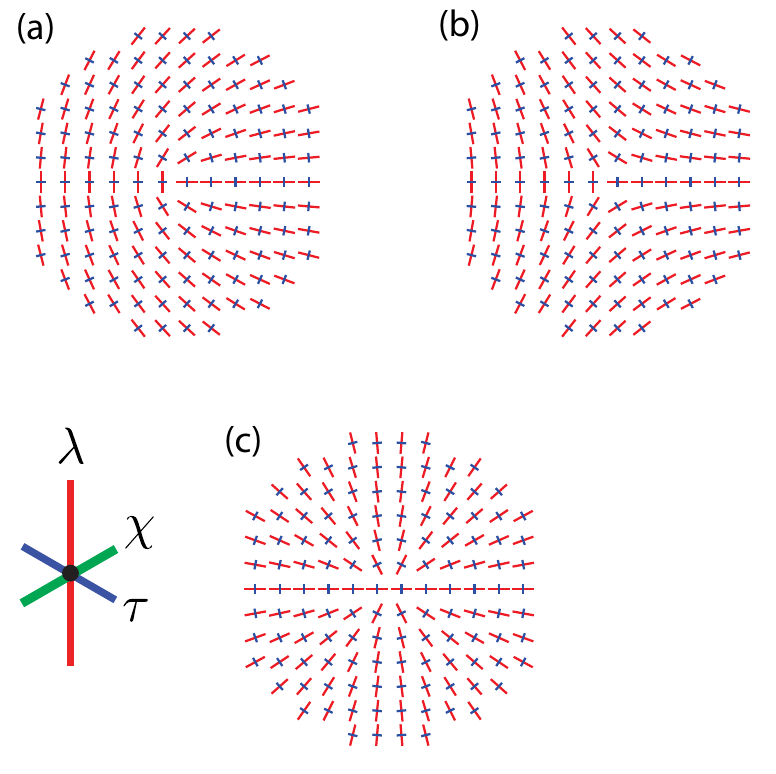}
        \caption{
        (a) $+\chi$ representative of the $C_\chi$ class, forming a $1/2$ disclination in both the $\lambda$ and $\tau$ fields and nonsingular in $\chi$. (b) $-\chi$ representative of the $C_\chi$ class, forming a $-1/2$ disclination in $\lambda$ and $\tau$ and nonsingular in $\chi$ (same as Fig.~\ref{fig:brickmap}(h)).
        (c) One of the three representatives of the $-1$ class of defects. Note that unlike in 3D uniaxial nematics, such defect configuration is topologically stable.}
        \label{fig:defects}
\end{figure}

A subtler consequence of having a  non-commutative fundamental group is that one-dimensional defects are more properly classified by the associated conjugacy classes~\cite{hatcher2002algebraic, mermin_topological_1979}. These classes are represented pictorially in Table~\ref{tab:group_table}. Biaxial systems have 5 conjugacy classes $Q_8=\{1\}\cup\{-1\}\cup C_\lambda\cup C_\tau\cup C_\chi$ where $C_\lambda=\{\pm \lambda\}$, $C_\tau=\{\pm \tau\}$, $C_\chi=\{\pm \chi\}$.  One can interpret the defect class $C_{\lambda/\tau/\chi}$ as a defect texture in which there is a singularity in all directions except the direction corresponding to $\lambda/\tau/\chi$. Examples of defects of class $C_{\chi}$ (singular in both $\lambda$ and $\tau$ but non-singular in $\chi$), as well as a non-removable $-1$ disclination, are shown in Fig.~\ref{fig:defects}~\footnote{\label{fn1} Note that Fig.~\ref{fig:defects}(a) and Fig.~\ref{fig:defects}(b) can be continuously interchanged by a local $\pi$ rotation of each triad about the horizontal axis since they belong to the same conjugacy class. They are, however, topologically distinguishable as group elements by their winding numbers - they can morph one to the other by circumnavigating around a third defect, as shown in Fig.~\ref{fig:path_dep}}.

\section{Geometry of biaxial and chiral nematic defects}

As noted, the order parameter space spanned by the chiral triad comprised of the molecular axis, $\blambda$, and the helical axis, $\bchi$, shares the same structure as the biaxial triad and thus has the same  ground state topology~\cite{volovik1977superfluid,michel1980symmetry, pollard_contact_2023}. An alternative description of chiral nematics has been recently studied experimentally in molecular-colloidal chiral hybrid systems ~\cite{wu2023unavoidable, mundoor2018}.  Such systems exhibit biaxiality that depends on the colloidal concentration. It was found that, in chiral nematic hosts, biaxiality persists even at low concentrations   ~\cite{wu2023unavoidable}. Moreover, the existence of a mapping from the elastic properties of chiral nematics to the biaxial free energy, described above for conventional chiral nematics and in  ~\cite{ senyuk2021nematoelasticity} for hybrid molecular-colloidal chiral nematics, allows one to analyze chiral nematic defects as a biaxial system. Here we treat the biaxial interpretation of CLCs as an approximation that is exact on the ground state manifold.

\section{Non-Abelian Signatures}
\label{sec:NonAbelian}
\subsection{Path Dependence}

\begin{figure}
        \centering 
        \includegraphics[width=.9\columnwidth]{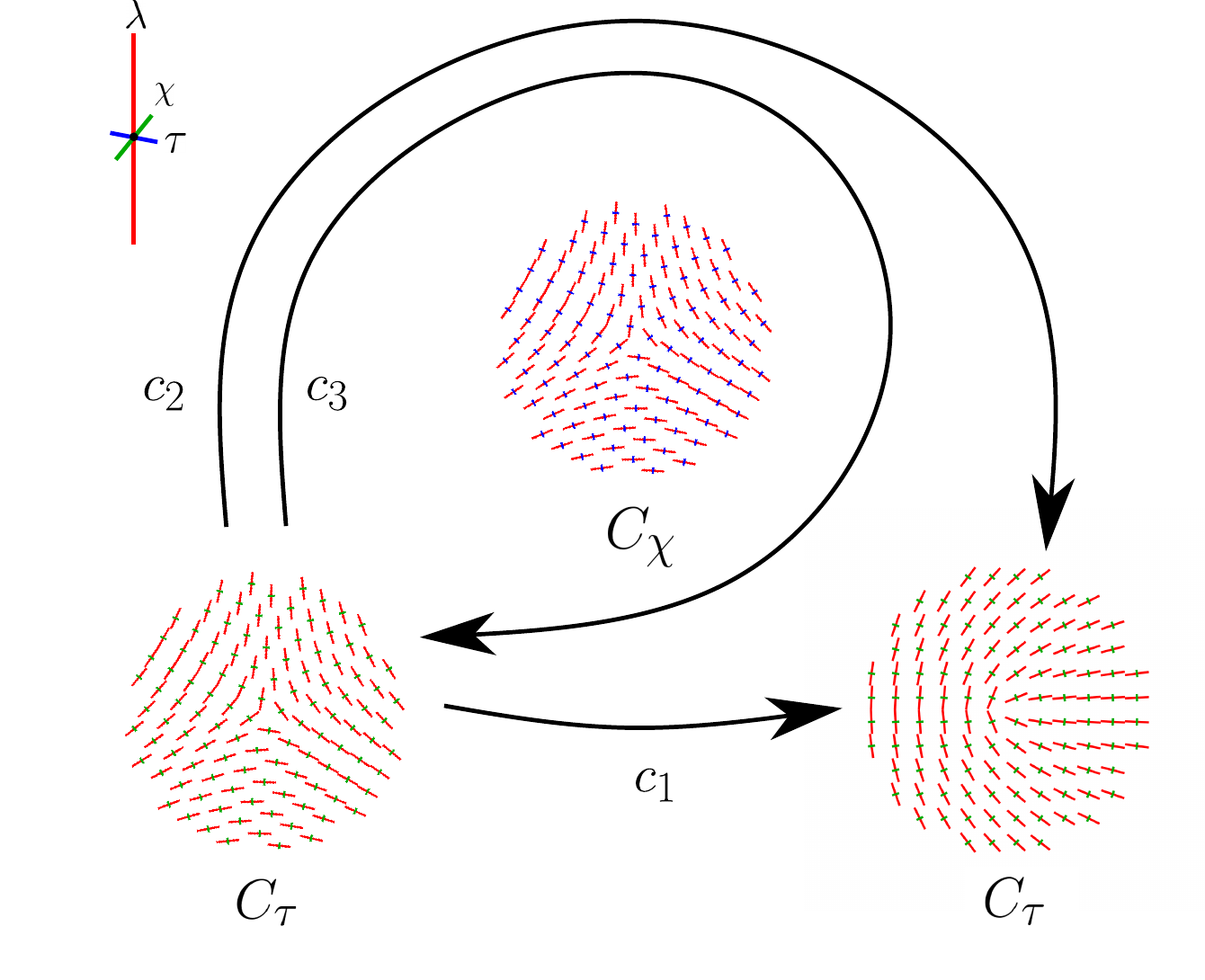}
        \caption{
        Fusion of two defects belonging to the same class, showing the path dependence of the resulting defect. The top path braids the defect around a third defect belonging to a different class, $C_\chi$, which continuously rotates the director structure of the defect without changing it's conjugacy class resulting in a $-1$ defect. The bottom path avoids the third defect, resulting in annihilation.}
        \label{fig:path_dep}
\end{figure}

The loss of commutativity for a non-Abelian fundamental group of an order parameter space has as its main consequence the loss of path independence in the composition of loops~\cite{mermin_topological_1979, kleman2003soft}. Mathematically, the non-commuting elements only allow one to relate two fundamental groups defined at different points up to inner automorphisms generated by a path connecting the two base points~\cite{hatcher2002algebraic}.

This path dependence manifests itself physically as an ambiguity in the combination of two defects -- one must know the history of the paths the defects have taken to combine. As shown in Table~\ref{tab:conj_table}, the combination of two defects of the same class, say $C_\tau$, leads to two different results depending on the path taken. Fig.~\ref{fig:path_dep} is an example of two such paths. In the presence of a third defect, such as $C_\chi$, we can form two non-homotopic curves, $c_1$ and $c_2$, that  are possible path histories. The path given by $c_1$ avoids the third defect and leads to the two defects annihilating, which is algebraically equivalent to a $+1$. The path given by $c_2$ is equivalent to the composition $c_1\circ c_3$, which braids the left defect around the top defect along $c_3$ and then takes $c_1$ to combine the defects. This total path leads to a $-1$ defect line, meaning that the act of braiding changes the composition of the defect without changing its topological classification under the lens of the fundamental group.  

\subsection{Entanglement of Defects}

\begin{figure*}
        \centering 
        \includegraphics[width=\textwidth]{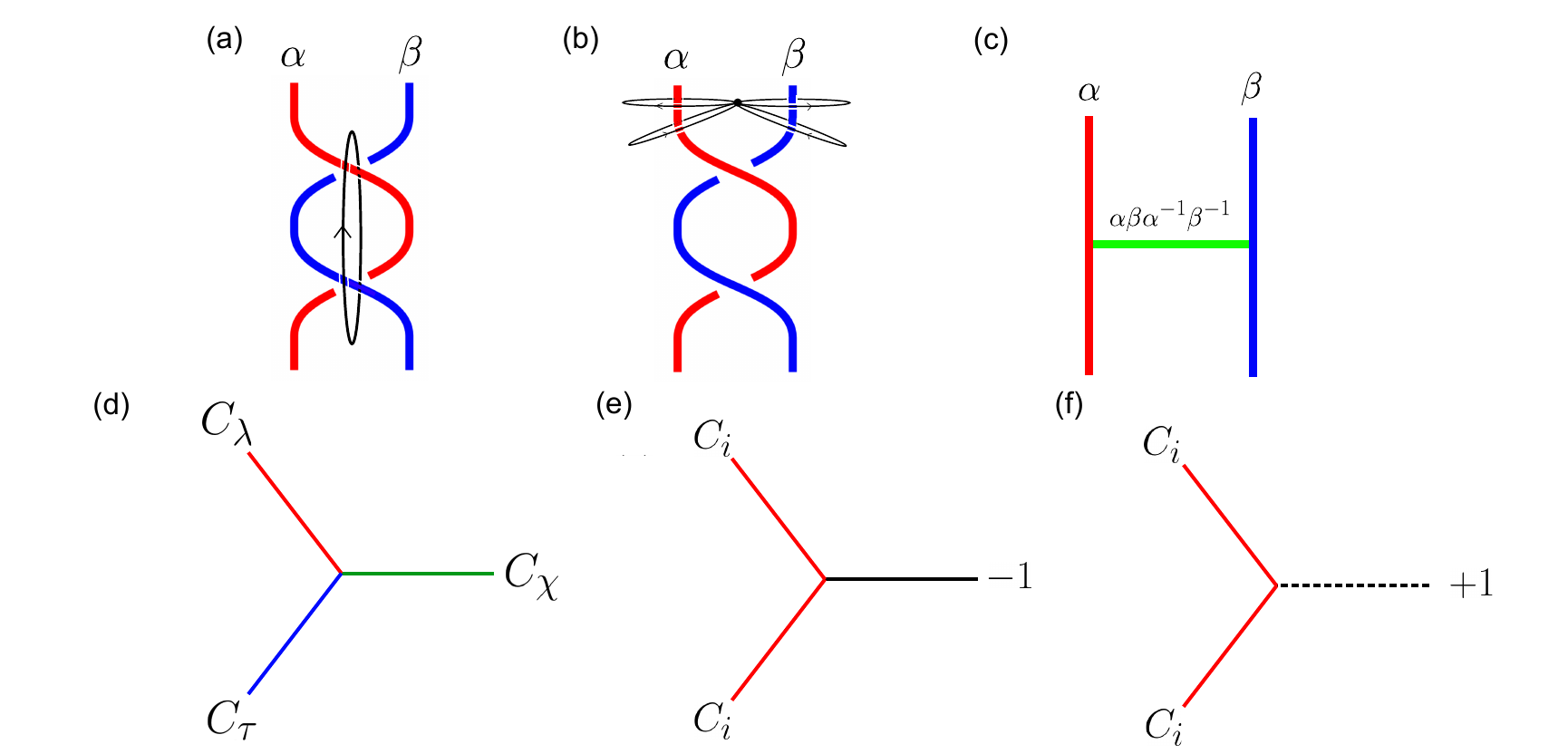}
        \caption{(a) Braided structure formed by two defect lines of conjugacy classes $\alpha$ and $\beta$. The black loop classifies the homotopy class of the braided crossing. (b) The black loop can be isotopically deformed to the equivalent set of 4 loops shown which reveals that pulling the lines apart results in a third defect of charge given by the commutator of the two initial defects, $[\alpha,\beta]=\alpha\beta\alpha^{-1}\beta^{-1}$ as shown in (c). (d-f) Three classes of trivalent junctions formed from the quaternion algebra. (d) Junction consisting of three disclinations belonging to three distinct conjugacy classes (e-f) Junctions that represent the path-dependent result of combining two disclinations in the same conjugacy class, $C_\lambda^2=+1\oplus -1$, respectively. Note that similar junctions as (e) and (f) also exist for $C_\tau$ and $C_\chi$ classes. }
        \label{fig:braid}
\end{figure*}

Arguably of more importance is the result of braiding two non-Abelian defects around each other. A loop drawn around the crossing can be deformed up to homotopy to show that the configuration is equivalent to the commutator of the two defect classes~\cite{poenaru_crossing_1977, mermin_topological_1979, alexander2022entanglements}. 

To see this, suppose we take two defects $\alpha,\beta\in Q_8$ and braid them around each other as shown in Fig.~\ref{fig:braid}(a). In order to figure out the topological classification of the effective defect generated by the braid we may draw a loop, $c$, around the middle braid (see Fig.~\ref{fig:braid}(a)). We may continuously deform $c$ in such a way that we can form the collection of loops shown in Fig.~\ref{fig:braid}(b). We recall that a loop around defect lines defines the topological charge of the defect, so each of these loops signify different factors of the corresponding defect charge. In other words, if we were to pull apart the braid, the effective defect charge formed in the middle can be read off from the loops in Fig.~\ref{fig:braid}(b) and it corresponds to the commutator 
    \begin{equation}
        [\alpha,\beta] = \alpha\beta\alpha^{-1}\beta^{-1}.
    \end{equation} 
Due to the structure of the quaternion group, this commutator can only result in either $\pm1$ (see Table~\ref{tab:conj_table}). Technically the commutator subgroup of $Q_8$ is $\mathbb{Z}_2$. The trivial result occurs when $\alpha$ and $\beta$ belong to the same conjugacy class or one of the $\pm 1$ classes. On the other hand, when the two lines belong to different conjugacy classes, we obtain the non-trivial result, $-1$. This means there are non-trivial entangled structures that are topologically stable to external fluctuations.  


Considering the chiral-to-biaxial nematic mapping described above, we can now pursue an experimental verification of this key classical prediction~\cite{ mermin_topological_1979}.



\subsection{Junctions} \label{subsec:Junctions}
Another important consequence of having a quaternionic fundamental group is the existence of non-trivial trivalent junctions. The algebra of the quaternion conjugacy classes states that combining two defects of two different classes ($\ne\pm1$) results in a defect in the third. This is simply a restatement of the usual cyclic identities of the quaternions, all contained within $ijk=-1$. These trivalent junctions provide elementary building blocks from which one can construct more complex structures, such as trivalent networks and lattices. 

Table~\ref{tab:conj_table} shows the types of junctions possible. Quaternion elements have the property that $i^2=j^2=k^2=-1$, which in terms of conjugacy classes reads $C_\lambda^2=C_\tau^2=C_\chi^2=+1\oplus -1$, where we drop the degeneracy of 2. This leads to junctions of the type shown in Fig.~\ref{fig:braid}(c), where two defects of the same conjugacy class meet with a $-1$ line. There is also a  trivial junction joining two defects of the same class to the identity element.


\subsection{Network structures}

\begin{figure}
    \includegraphics[width=\columnwidth]{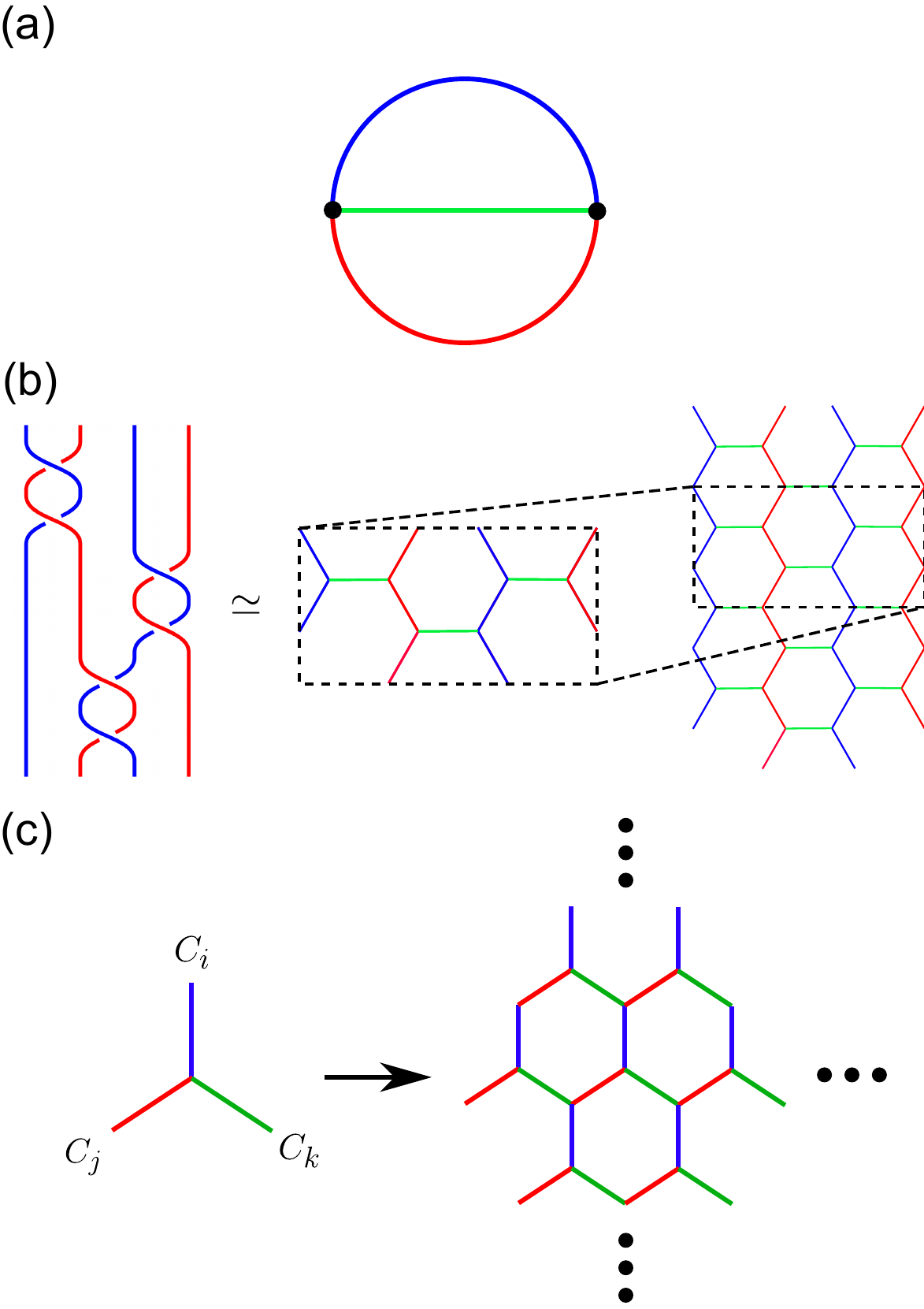}
    \caption{(a) Simplest possible connected structure made from connecting all the free ends from the junction in (c) together. Note that such structure can be obtained by "fusing" closed loops of defect lines belonging to distinct classes, say i and j. (b) Braided structure of four defect lines belonging to two different conjugacy classes gives rise to a topologically equivalent network ``unit cell'' where the green line belongs to the $-1$ class. This unit cell can be repeated to form a junction network (right). (c) A stable junction of three defect lines belonging to separate conjugacy classes also generates a unit cell for more complex lattice structures (right). }
    \label{fig:braid network}
\end{figure}

Among the possible complex structures there are non-trivial links and networks (Fig.~\ref{fig:braid network}). The simplest possible network can be realized minimally with two junctions. Take, for example, a $(\lambda,\tau,\chi)$-junction in Fig.~\ref{fig:braid}(d). One can take two legs and connect them on the third leg, creating a closed two-junction structure (see Fig.~\ref{fig:braid network}(a)). 

Here we focus on more complex structures. One such structure can be realized from the entanglement viewpoint as follows: considering several defect lines of alternating defect classes, we then perform the braid operation shown in the left hand side of Fig.~\ref{fig:braid network}(b). This structure is topologically equivalent to the set of junctions shown in the right of Fig.~\ref{fig:braid network}(b). Such a structure can then be repeated to form an arbitrarily large hexagonal network, which can be seen as either a network or a braid. Similarly, one can create a repeating network of $(\lambda,\tau,\chi)$-junctions as shown in Fig.~\ref{fig:braid network}(c). In practice, implementation of such networks will have the topology shown in Fig.~\ref{fig:braid network}, although the geometry may not the exactly hexagonal due to energetic differences in the elastic tension of each defect line following from the precise values of the elastic constants in Eq.~\ref{FrankOseenE}. 
    
\section{Methods}

\subsection{Simulations}
\label{subsec:sim_methods}

\subsubsection{Numerical simulations}

We model the CLC equilibrium structures by numerical minimization of the Landau-de Gennes expansion of bulk free energy density, including terms given by Eqs.(\ref{Qenergy},\ref{ChiralTerm}) and the thermotropic term governing the isotropic-nematic phase behavior:
    \begin{align}
        F_\mathrm{bulk} &= \int \mathrm{d}r^3 ~\bigg( 
        \frac{A}{2}\mathrm{Tr(\mathbf{Q}^2)} 
        + \frac{B}{3} \mathrm{Tr(\mathbf{Q}^3)} 
        + \frac{C}{4} \mathrm{Tr(\mathbf{Q}^2)}^2 \nonumber \\
        &+ \frac{D}{5}\mathrm{Tr(\mathbf{Q}^2)Tr(\mathbf{Q}^3)}
        + \frac{E}{6}\mathrm{Tr(\mathbf{Q}^2)}^3
        + \frac{F}{6}\mathrm{Tr(\mathbf{Q}^3)}^2
        \nonumber \\
        &+ f_{\mathrm{elastic}}+f_{\mathrm{chiral}} \bigg) \ ,
    \end{align} 
where $Q$ is the tensorial order parameter; $A$, $B$, and $C$ are parameters related to the isotropic-uniaxial phase behavior of CLCs and $D$, $E$ and $F$ determine the uniaxial-biaxial nematic transition \cite{gramsbergen1986landau,vissenberg1997generalized,allender2008landau}. Here we only consider the nematic-isotropic transition for a chiral nematic system and keep up to 4th-order terms.
Without higher order (5th and 6th) terms and with biaxiality dominated by $f_{\mathrm{chiral}}$, we have a precise mapping of chiral and biaxial director structures shown in Fig.~\ref{fig:brickmap}.
The biaxial directors can still be well defined and the topology of the D2CLC is still robustly reconstructed when also considering the additional contribution to biaxiality from non-chiral energies, including higher order (5th and 6th) terms.

In some cases with disclinations connected to the boundary of the numerical volume, surface boundary conditions $\mathbf{Q}^0$  for $\mathbf{Q}$ are added to the calculation of the total energy via
    \begin{equation}
        F_\mathrm{surface} = \int \mathrm{d}r^2 ~\frac{W}{2} (\mathbf{Q}-\mathbf{Q}^0)^2 \ ,
    \end{equation}
with $W$ being a surface anchoring coefficient.
Equilibrium structures are found based on a gradient descent method with a finite difference mesh~\cite{sussman2019fast}, using a home-built Matlab program. 
The director fields and the uniaxial and biaxial order parameters are obtained by identifying the eigenvalues $\lambda_i$ and eigenvectors $\mathbf{v}_i$ of the tensor~\ref{Qtensor} at each grid point~\cite{mucci2016landau}:
    \begin{align}
        S &= \lambda_1-\lambda_3 \nonumber \\
        \mathbf{n} &= \mathbf{v}_1 \nonumber \\
        T &= \lambda_2-\lambda_3 \nonumber \\
        \mathbf{m} &= \mathbf{v}_2 \ ,
        \label{Qeigen}
    \end{align}
with $\lambda_1>\lambda_2>\lambda_3$ and corresponding $\mathbf{v}_i, i=1-3$ being the eigenpairs of the energy-minimizing $\mathbf{Q}$. 
The third director is then $\mathbf{l}=\mathbf{n} \times \mathbf{m}$. In the approach described here, our biaxiality analysis and interpretation of disclinations is not hindered by the emergence of vanishing biaxiality or reverse-twisting at point defects \cite{ackerman2017diversity}, since the biaxial topology prescribed by chirality and arising due to other elastic deformations is consistent throughout the volume. This is because it is energetically costly to fully unwind the twist structure to a uniformly aligned state, even with such small biaxiality $T$.

In the examples illustrated in this work, the following parameter values were used:  $A=-1.72 \times 10^5 \jmt$, $B=-2.12 \times 10^6 \jmt$, $C=1.73 \times 10^6 \jmt$, $\gamma_1=7.4 \pN$, $\gamma_2=12 \pN$, $\gamma_4=-0.0017 \textrm{N}/\textrm{m}$, $\gamma_6=11.9 \pN$, and $W=10^{-3} ~\textrm{J}/\textrm{m}^2$. These constants give an equilibrium $S\approx 0.6$ and $T\approx 0.001$. Note that a negative $\gamma_4$ represents left-handed chirality in our numerical model. 

\subsubsection{Numerical visualization}
The analysis and visualization of disclination positions were performed in Matlab by rendering the smoothness of the director fields. For instance, a $\lambda$ disclination is represented by discontinuity in the $\bchi$ and $\btau$ director fields but not in the $\blambda$ field \cite{kleman1969lignes}.  $\tau$ disclinations, likewise, are revealed by finding regions with a continuous $\btau$ field but singularities in the $\blambda$ and $\bchi$ fields. In practice, we quantified the smoothness of a director field based on the inner product of the directors at neighboring grid points. 

Another more conceptually straightforward, but more computationally intensive approach, is to calculate the actual topological winding numbers in the three director fields for all grid points. A $\lambda$ disclination is then identified by vanishing winding number in $\blambda$ but nonzero ($\pm\frac{1}{2}$) winding in the other two director fields. The contour regions of director smoothness, or winding number, are plotted for each field individually and then superimposed. The two methods give consistent results. 

For defects in which the helical axis director $\bchi$ is not uniquely defined \cite{Efrati_2014}, we performed the visualization based instead on the matching biaxial directors $(\mathbf{n},\mathbf{m},\mathbf{l})$ (see \ref{sec:BiaxialElasticity}), obtained from tensor diagonalization (Eq.~\ref{Qeigen}). In some cases we manually picked the combination of regions and one of the two director sets to circumvent numerical artifacts.

\subsection{Experimental techniques}

\subsubsection{Sample preparation} 
The CLC cells in our work were made of two glass slides or coverslips, both chemically treated to have specific designed surface anchoring, determined by the  boundary conditions for the LC molecular directors. Specifically, 1.0 wt\% (weight percentage) of poly(vinyl alcohol) (PVA, Sigma-Aldrich) in water was used to generate a unidirectional parallel configuration for the CLC directors, and an azobenzene dye SD1 1.0 wt\% in dimethylformamide was applied for photo-alignment~\cite{meng2023topological,chigrinov2005photo}. Chemical solutions were evenly spread on the glass surfaces by spin-coating at 700 rpm for 15s then 3,000 rpm for 45s, followed by heating at 100$^\circ$C for 1 min to thoroughly evaporate solvents. For PVA-coated surfaces, the anchoring direction was defined by gently rubbing a piece of cloth against the surface along the desired anchoring direction. The boundary conditions for SD1-treated glasses, on the other hand, were designed using photo-alignment techniques (detailed below)~\cite{meng2023topological,chigrinov2005photo} and surface patterns with topological point defects and domain walls were imprinted to facilitate the formation of disclinations.

The gap distance between the two glass pieces was set by silica spacer spheres (with diameters ranging from 7 to 40 $\um$, from Thermo Fisher). The CLC sandwiched between them was prepared by doping cholesterol pelargonate (Sigma-Aldrich) into 4-Cyano-4'-pentylbiphenyl (5CB, EM Chemicals) to form a left-handed CLC and subsequently used to fill the glass cells. The cholesteric pitch $p$, controlled by the doping concentration, ranged from 2 to 10 $\um$. 

\subsubsection{Photo-patterning of the confining surface}
To precisely control the types and locations of defects, we photo-patterned the SD1-coated surfaces with topologically nontrivial boundary conditions.
During the patterning, the photo-responsive azobenzene dye was locally reoriented by polarized blue light to render it perpendicular to the incident polarization in the controlled illumination area.
We performed the illumination with a commercial microdisplay and generated configurations based on identifying regions with the same director orientations in the pre-designed director boundary conditions \cite{martinez2011large}.
The optical alignment involved a waveplate and a polarizer to adjust the polarization of the illumination light and a 4× objective (numerical aperture, NA 0.13) to focus the light on our sample glass surface. The detailed optical setup for photopatterning can be found in Refs.~\cite{meng2023topological,martinez2011large}.
CLCs were subsequently introduced into the glass cell, after which we found the reorientation of azobenzene dye to be negligible.

To observe non-Abelian defects, a charge-coupled device camera (PointGrey) mounted on inverted microscopes (Olympus, IX-83) with 4× objective (Olympus, NA 0.13) was used for optical microscopy, such as bright-field imaging. Polarizers and a phase ring were inserted for polarizing optical microscopy and phase contrast microscopy, respectively. 
Three-photon excited fluorescence polarizing microscopy (3PEF-PM)~\cite{lee2010multimodal} was also carried out to verify the molecular director $\bn$ configuration around defects. In brief, the three-photon excitation of the 5CB molecules involved the incident light generated from a Ti:sapphire pulse laser (Chameleon Ultra II, Coherent) operating at 870~nm wavelength passing through a linear polarizer, and the fluorescence signal was epi-collected using a 60× objective (Olympus, NA 1.35) and magnified by a photomultiplier tube (H5784-20, Hamamatsu). 
In the three-photon process, the fluorescence intensity scales as $\cos^6\theta_{\mathrm{n}}$, where the angle $\theta_{\mathrm{n}}$ is the angle made by the 5CB molecular director $\blambda$ with the polarization of the excitation light.
The microscopy was thus used to probe the director alignment and numerical simulations of 3PEF-PM images were performed accordingly.

\section{Results}

\begin{figure}
        \centering 
        \includegraphics[width=\columnwidth]{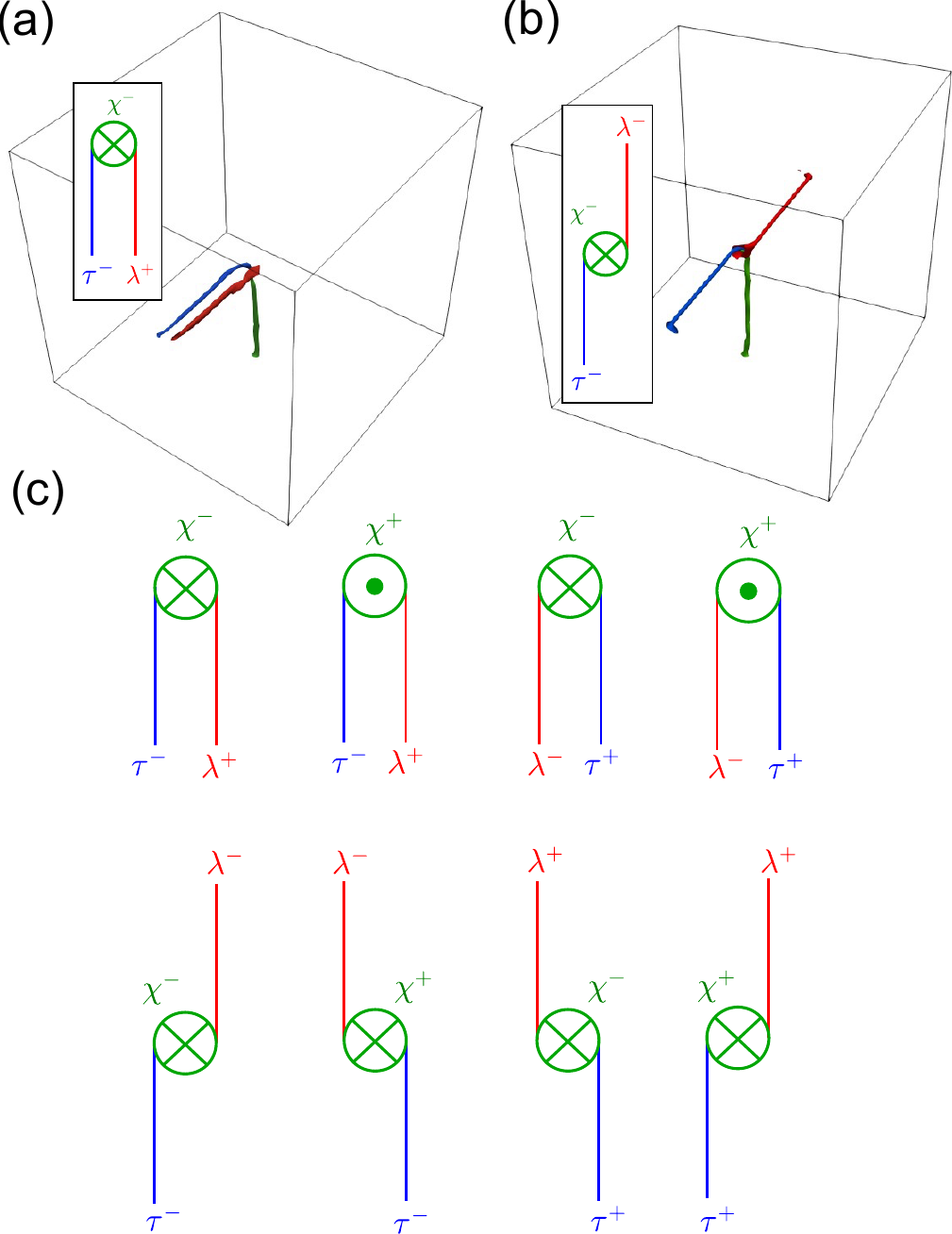}
        \caption{(a-b) Numerically simulated $\chi/\lambda/\tau$ 3-fold junctions for the cases (a) $\tau^{-}\lambda^{+}\longleftrightarrow\chi^{-}$ and (b) $\tau^{-}\longleftrightarrow\chi^{-}\lambda^{-}$ with their respective diagrams shown.  The results are obtained from the minimization of the Landau-de Gennes energy (Methods). (c) Schematic representations of 8 geometries of junctions obtained in numerical computation of a left-handed chiral nematic LC. Here, the direction of the $\chi$ disclination is represented by a cross and a dot for into and out of the page, respectively.}
        \label{fig:junction_sim}
\end{figure}

\begin{figure}
        \centering 
        \includegraphics[width=\columnwidth]{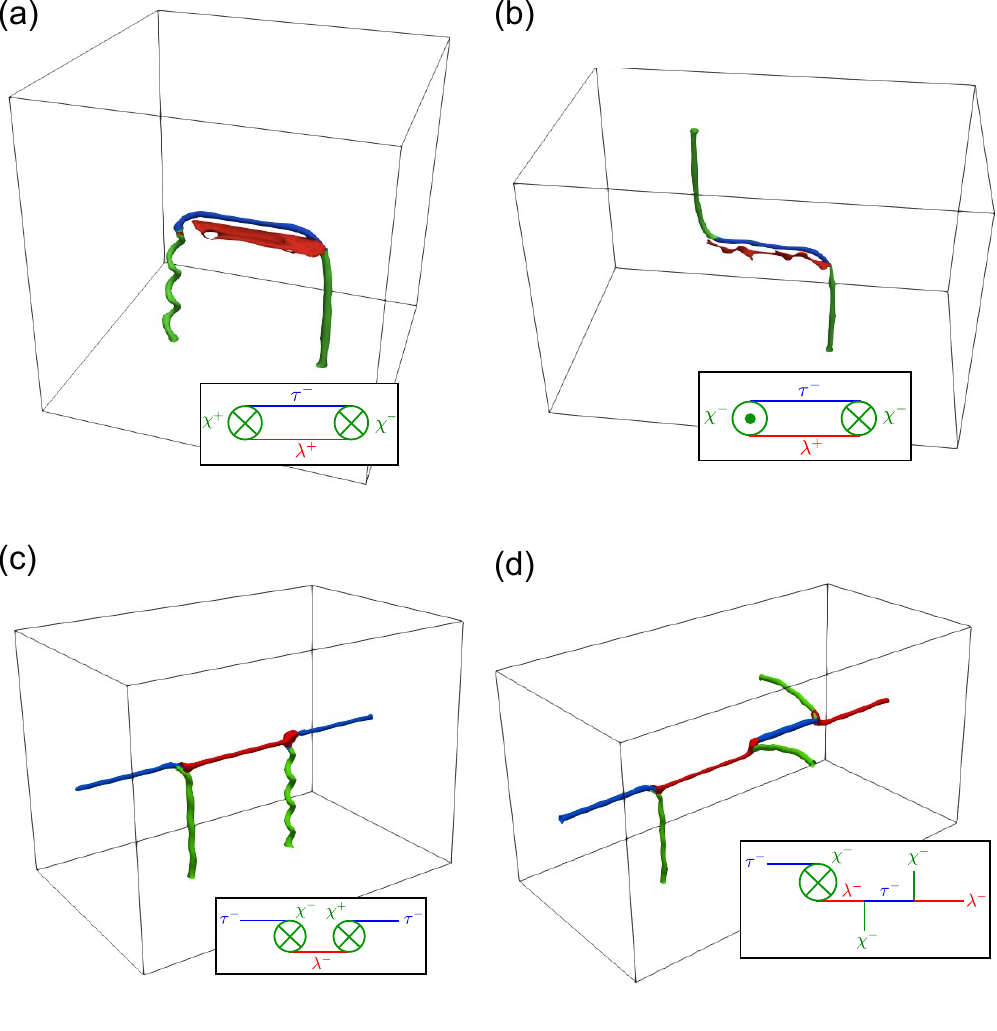}
        \caption{Examples of numerically modeled disclination structures based on $(\chi,\lambda,\tau)$ junctions with their corresponding diagrammatic representations. (a-b) Two examples of joined elementary trijunctions with $\chi$ disclination ending on surfaces. (c-d) Two examples of composite junction series comprised of two and three elementary junctions, respectively, can be extended to larger disclination networks.
        }
        \label{fig:junction_sim2}
\end{figure}

\begin{figure*}
        \centering 
        \includegraphics[width=\textwidth]{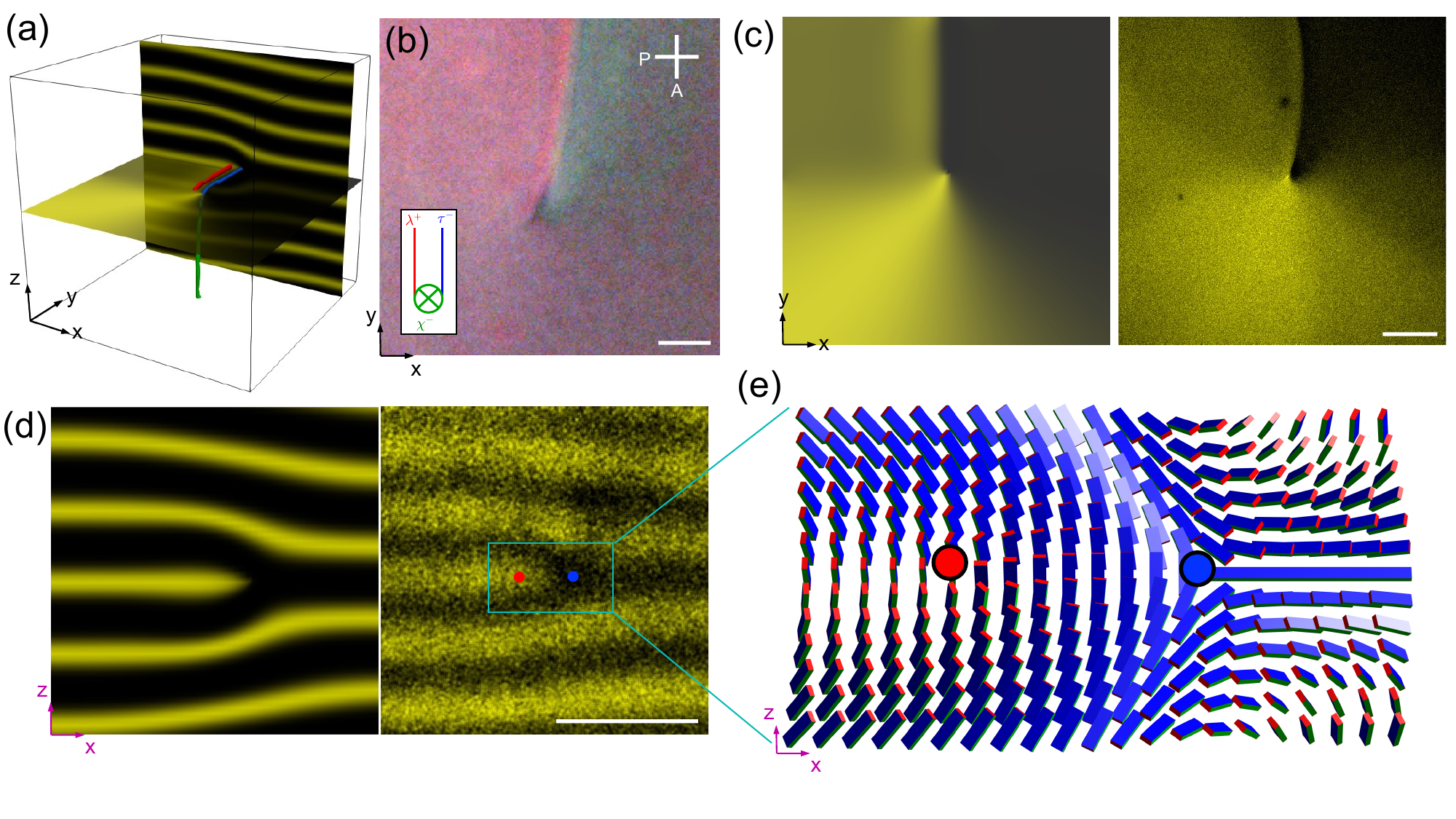}
        \caption{
        (a) A defect junction visualized in computer simulations. (b) Experimental top view of such defect junction obtained with polarizing optical microscopy (POM). Diagrammatic representation of the junction; orientations of polarizer and analyzer are marked. (c,d) Simulated (left) and experimental (right) 3-photon excitation fluorescence polarizing microscopy (3PEF-PM) images along or across the $\lambda$-$\tau$ disclination pair. The slice planes are also shown in (a). (e) The zoomed-in brick visualization where brick edges represent biaxial director fields (from long to short: $\bn$, $\mathbf{m}$, $\mathbf{l}$ respectively). The 3PEF-PM signal is strongest when molecular director $\bn$ is aligned along the $y$ direction (see Methods). The junction structure is also described as a dislocation in systems viewed as chiral nematic quasi-layers. All scale bars are 5 $\um$. 
        }
        \label{fig:junction_exp}
\end{figure*}

\begin{figure*}
        \centering 
        \includegraphics[width=\textwidth]{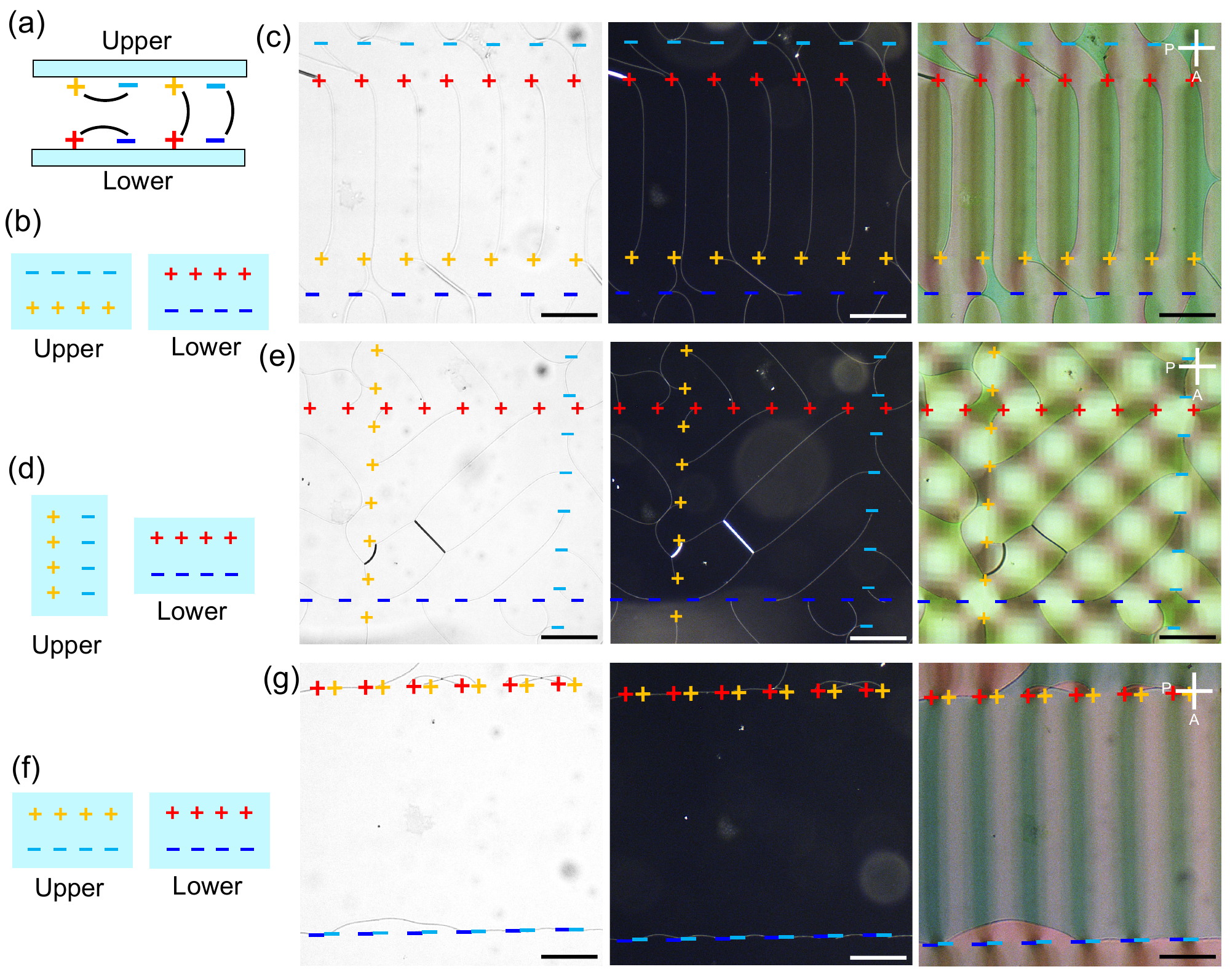}
        \caption{(a) Topologically selected defect connections (black curves) based on the junction geometries in Fig.~\ref{fig:junction_sim}(c). The isolated, uninterfered defect lines only connect two defects on the same surface or with the same sign, but not both. The ``+'' and ``-'' signs represent surface point defects with winding number $\pm1/2$ respectively. Defects on different surfaces are distinguished by colors.  (b,d,f) Schematics representing the upper and lower boundary conditions for LC directors with half-integer defects. (c,e,g) Corresponding connections between surface defects when upper and lower glass plates are overlaid. 
        Cholesteric pitch $p=7.4 \um$, cell thickness $d=50 \um$, and scale bars are $300 \um$.
        }
        \label{fig:defect_connect}
\end{figure*}

\begin{figure*}
        \centering 
        \includegraphics[width=\textwidth]{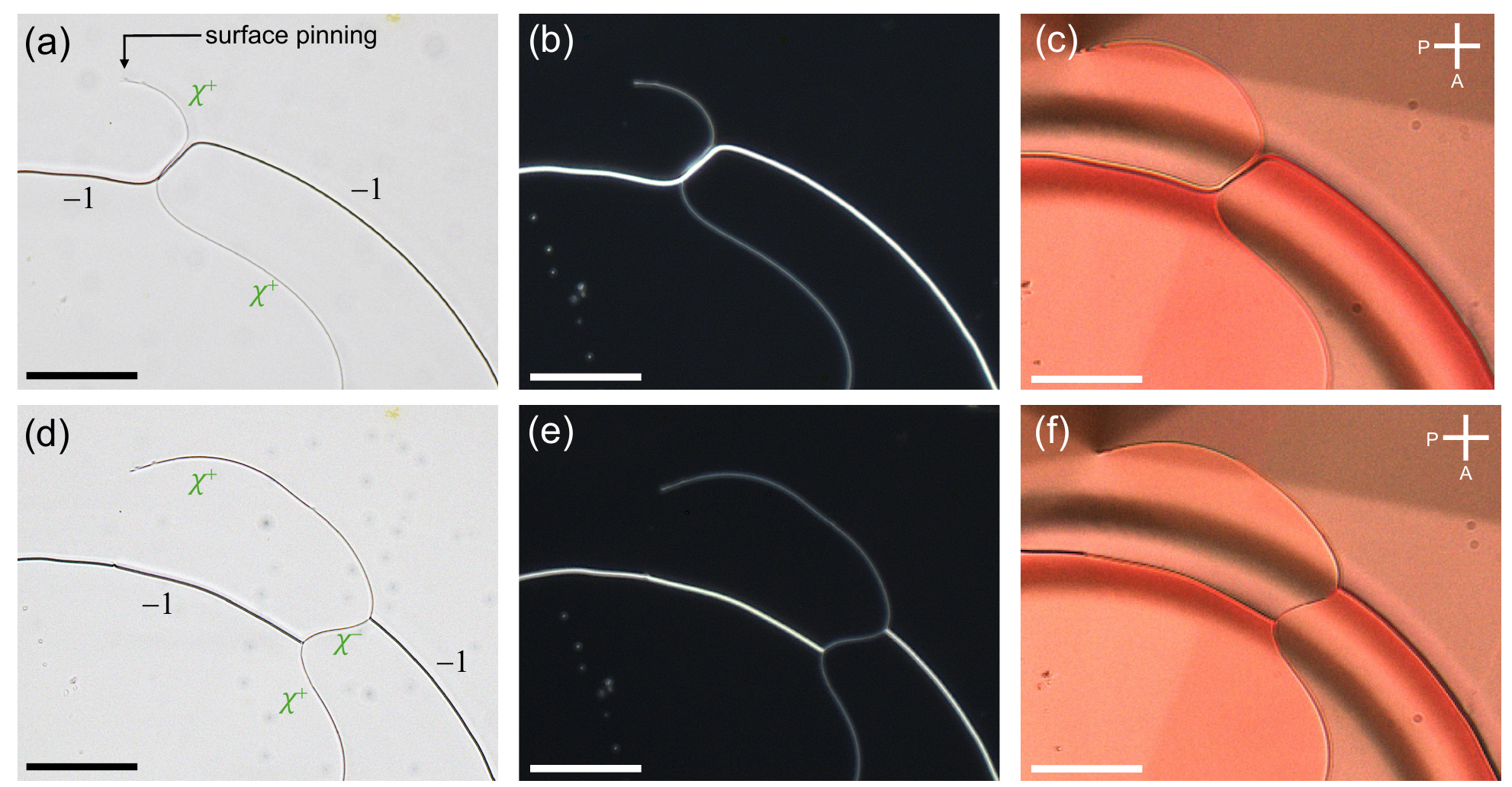}
        \caption{ (a-c) Cross-over of a -1 and a $\chi^{+}$ disclination imaged using (a) bright field, (b) phase contrast, and (c) polarized optical microscopy (POM), respectively. The $\chi^{+}$ line ends at a surface pinning point defect on the upper boundary of the CLC sample (see Methods). (d-f) Two topologically stable 3-fold junctions formed from merging the cross-over in (a-c) accompanied by the generation of $\chi^{-}$. Orientations of polarizers are marked in (c,f), cholesteric pitch $p=7.4 \um$, cell thickness $d=50 \um$ and scale bars are 100 $\um$. The displacement of the junctions is due to the fluidity in our soft matter system. 
        }
        
        \label{fig:line_merge}
\end{figure*}

\begin{figure*}
        \centering 
        \includegraphics[width=\textwidth]{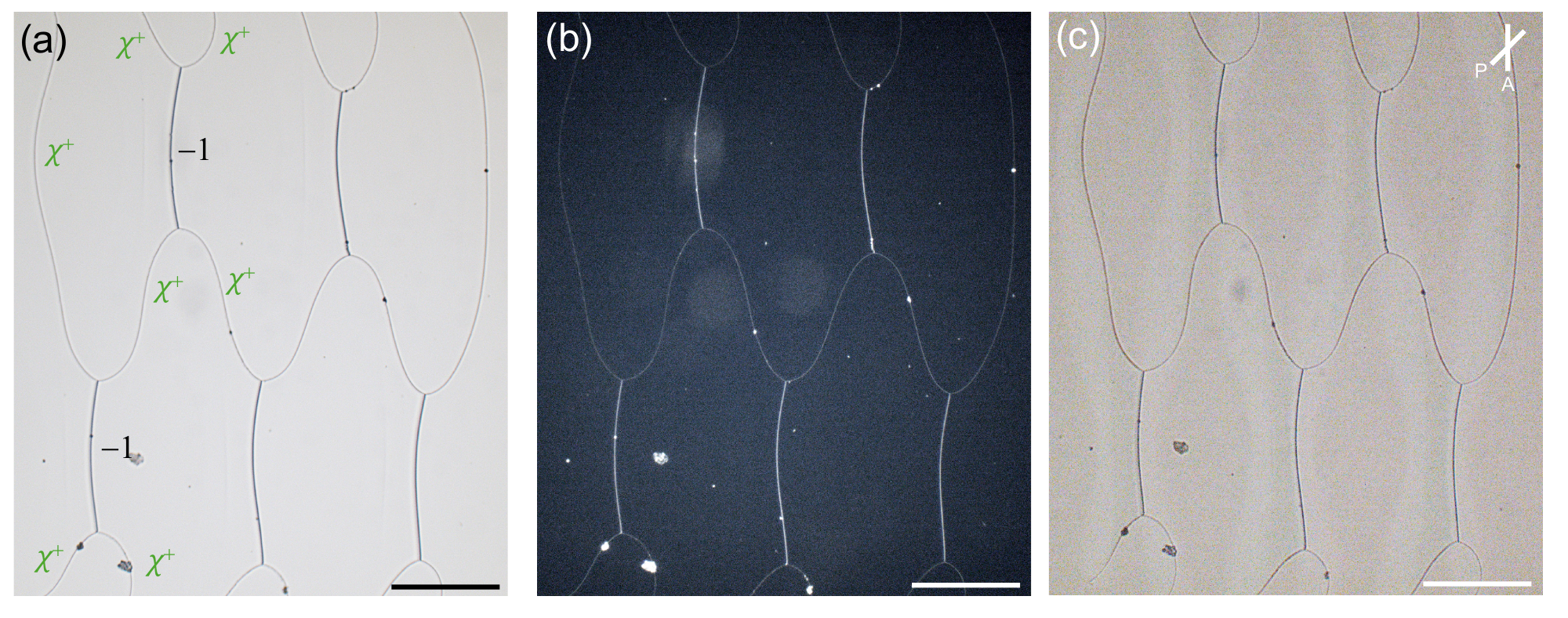}
        \caption{(a-c) Experimental realization of a stable disclination network based on the ($\chi,\chi,-1$) junctions imaged using (a) bright field, (b) phase contrast, and (c) polarized optical microscopy, respectively. Cholesteric pitch $p=7.4 \um$, cell thickness $d=50 \um$, and scale bars are 300 $\um$. 
        }
        \label{fig:network}
\end{figure*}




The quaternion algebra governing biaxial defects contains three classes of three-defect junction (see Fig.~\ref{fig:braid}(d)). Here, we describe observations and analysis of such structures in both computer simulations and experiments. While experimentally various junctions are observed in unconstrained sample geometries, they are typically uncontrolled transient features of polydomain samples. In contrast, our surface photopatterning approach allows us to pin the end points of these defect lines so that their structure can be thoroughly analyzed. 

\subsection{\texorpdfstring{$(\lambda,\tau,\chi)$}{} Junctions}

We first simulate a chiral nematic system with the methods laid out in Section~\ref{subsec:sim_methods}. 
Fig.~\ref{fig:junction_sim} shows two different simulated $(\lambda,\tau,\chi)$ junctions corresponding to the processes (a) $\tau^{-}\lambda^{+}\longleftrightarrow\chi^{-}$ and (b) $\tau^{-}\longleftrightarrow\chi^{-}\lambda^{-}$. These disclination junctions are not only topologically stable but also energetically robust against thermal fluctuations with the chosen numerical parameters.
Fig.~\ref{fig:junction_sim}(c) shows the complete set of junction types that are numerically observed and energetically relaxed for a left-handed CLC. Though one would expect that defects are free to deform to all the topologically equivalent structures conserved in such a soft fluid system, we note that the energy-minimizing junction structures obtained from computation inherit chirality from the underlying molecular chiral nematic host medium. The mirror-images of these junctions were found numerically to exist in right-handed CLCs, thus providing an extra control knob for chiral biaxial systems. In addition the structure of a bound pair of parallel $\lambda$ and $\tau$ disclinations (side-by-side) can be found in Fig.~\ref{fig:junction_sim}(c, top row) with notable energetic stability. In Section~\ref{sec:boundstate} we argue that the chirality of the system provides the elastic tension required for the formation of stable bound states of defects.

The three-fold junctions discussed so far consist of one $\lambda$, one $\chi$, and one $\tau$ disclination, that is, all of the class shown in (Fig.~\ref{fig:braid}(d)). Beyond the single-junction (elementary) structures, multi-junction strutures were also numerically stabilized by gluing junctions with matching topology and geometry. Fig.~\ref{fig:junction_sim2}(a,b), for example,  shows how a $\chi$ disclination splits into a parallel pair of $\lambda$ and $\tau$ defects, which then merge to form a $\chi$ again. Interestingly, numerical minimization of the free energy suggests that there is a preference for $\chi^{-}$ or $\chi^{+}$ depending on the relative orientation of the defects. In other words, energy-minimizing structures of multiple junctions are assembled from  connections of the geometries found in Fig.~\ref{fig:junction_sim}(c). As a result, we found that $\chi$ defects connected at the same side of the numerical volume tend to have the opposite signs (Fig.~\ref{fig:junction_sim2}(a)), and conversely $\chi$ lines with the same sign extend out in different directions as in Fig.~\ref{fig:junction_sim2}(b). This feature of the junction geometry is further confirmed experimentally below. 


3-fold junctions can also serve as the building blocks for multi-junction disclination networks, such as those shown in Fig.~\ref{fig:braid network}. By properly joining individual defect junctions, we numerically realized multi-junction networks that can be repeated to build infinitely large networks (Fig.~\ref{fig:junction_sim2}(d) illustrated a 3-junction network). 
Note that the equilibrium structure of $\chi^{+}$ in numerical simulations is a spiral-like trace of the disclination core due to the strong elastic anisotropy, $K_1, K_3 > K_2$ in Eq.~\ref{FrankOseenE}, while $\chi^{-}$ is straight and has lower elastic energy cost, as shown in Fig.~\ref{fig:junction_sim2}.  

Experimentally we also generated defect junctions and robustly-controlled disclination orientations by surface patterning techniques (see Methods). Fig.~\ref{fig:junction_exp} shows a tri-junction depicting the process $\chi^{-}\longleftrightarrow\lambda^{+}\tau^{-}$, with the $\chi$ disclination attached to the bottom surface defect. Since the $\chi$ is parallel to the viewing direction under the microscope (along $z$, Fig.~\ref{fig:junction_exp}(b)), it renders a dark spot where the $\lambda^{+}\tau^{-}$ defects terminate, characterized by the color contrast in polarizing optical microscopy (POM). Note that among simulated junctions Fig.~\ref{fig:junction_sim}(c) this form ($\chi^{-}\longleftrightarrow\lambda^{+}\tau^{-}$) is energetically favored by the the strong anisotropy of the system \cite{wu2022hopfions} and thus is frequently observed in numerics and experiments.

Furthermore, consistent numerical simulations of three-photon excitation fluorescence polarizing microscopy (3PEF-PM, see Methods) images were performed, which effectively reveal/confirm director configurations~\cite{lee2010multimodal}. We compared the director configuration around a defect junction in the whole three-dimensional (3D) volume shown in Fig.~\ref{fig:junction_exp}(a), which appears like a color contrast reaching an end under POM microscopy Fig.~\ref{fig:junction_exp}(b). Good agreement is found between computational and empirical results of the fluorescence signal, including the representative 2D slices in Fig.~\ref{fig:junction_exp}(c,d) parallel or perpendicular to the $\lambda\tau$ pair, or equivalently a $\chi$ defect. Fig.~\ref{fig:junction_exp}(e) shows the director configuration around a disclination pair, with bricks showing red faces indicating $\bn$ along $y$ direction and stronger 3PEF-PM signal in Fig.~\ref{fig:junction_exp}(d).This structure for $\bn$ is often found with mismatching helical layers and resembles a dislocation, as evident in Fig.~\ref{fig:junction_exp}(d).
The provides an in situ verification of the simulated defect structures as well as immediate identification of junction types. Since the vertical $\chi^-$ connects to the bottom surface with $-1/2$ winding number, we will label it as ``$-$'' for the experiments with multiple junctions demonstrated below.

As shown in Fig.~\ref{fig:junction_sim2}(a,b), the preferred geometries of defect connections in equilibrium are a combination of the defect junctions in Fig.~\ref{fig:junction_sim}(c). Simply put,  ``defects with same topological charge connect on opposite surfaces, while defects with opposite charge connect on the same surface'', as summarized schematically in Fig.~\ref{fig:defect_connect}(a). The case of opposite charge adds a handle to the boundary surface, providing a means of communicating between spatially separated points on the boundary via the bulk. The case of same charge defects allows one boundary surface to connect to a distant boundary via a line defect passing between them.  We then designed an experiment based on such selective defect connections (Fig.~\ref{fig:defect_connect}(b-g)). The preprinted surface defect with $+1/2$ or $-1/2$ winding number are placed in several relative positions. Fig.~\ref{fig:defect_connect}(b-c), for instance, shows that defects (each being $\lambda\tau\equiv\chi$) extend across the sample area to have larger disclination length instead of connecting to nearby surface defects. In contrast to a typical soft matter system where energy-costly defects would be shortened whenever possible, here the selective connections of disclinations are a result of the interplay between the non-Abelian biaxial topology and the geometry of non-Abelian defect junctions, on top of energetic considerations in a CLC. Here we focus on cleanly separated elementary defects and ignore those with distortions or with complex topology; these are thicker and brighter under the microscope~\cite{smalyukh2002three,wu2022hopfions}.
Furthermore, by simply rotating the upper glass substrate (Fig.~\ref{fig:defect_connect}(d-g)) we clearly observed disclination arrangements being ``same charge xor (i.e. exclusive or) same surface'' as summarized in Fig.~\ref{fig:defect_connect}(a), consistent with our interpretation of non-Abelian defect topology in CLCs.

\subsection{\texorpdfstring{$(\chi,\chi,-1)$}{} junction}

Another important type of tri-junction, beyond the class with one each of $\lambda$, $\chi$, and $\tau$ disclinations discussed above, is that resulting from the entanglement of two defects -- algebraically equivalent to the negation of a group element (Fig.~\ref{fig:braid}(e)). Experimental realizations of this junction are shown in Fig.~\ref{fig:line_merge}. For the crossover of a $\chi^{+}$ disclination with a $-1$ disclination see Fig.~\ref{fig:line_merge}(a-c). The individual creations of a $\chi$ line and a $-1$ defect on separated glass surfaces guaranteed the correct assignments of quaternion group elements (Methods). The crossing was then created by overlapping the two glass surfaces with their relative position carefully adjusted under the microscope. Near the defect crossing, the overall elastic energy is reduced by rendering the two defects parallel. Brightness and color contrast, however, allow for an unambiguous identification of the two distinct defect classes (Fig.~\ref{fig:line_merge}). By locally melting the CLC with laser tweezers near the defect crossing, and then allowing the system to reorient in a quench, we produced a set of two connected 3-fold junctions (Fig. ~\ref{fig:line_merge}(d-f)) of the $(\chi,\chi,-1)$ class. As evidenced by the microscopic characterization Fig.~\ref{fig:line_merge}(d-f), the two junctions are connected by a $\chi$ line corresponding to the defect algebra $\chi^{+}\times -1=\chi^{-}$, following Table.~\ref{tab:group_table}. We also assessed the topological stability of these junctions by repeated laser-tweezer-based heating and mechanical excitations: we found these $(\chi,\chi,-1)$ junctions to be unbreakable due to the topological protection imposed by intrinsic biaxiality. Furthermore, by recognizing this $-1$ defect as a combination of two parallel $\lambda$ disclinations, given its resemblance to a dislocation of one helical layer \cite{wu2022hopfions}, we can associate the $(\chi,\chi,-1)$ defect junction with a $\chi\chi\longleftrightarrow\lambda\lambda$ process.
We want to note that, however, such examples of $-1$ defect structure are different from those discussed in Ref.~\cite{beller_geometry_2014}, in which case the process of $\chi\chi\longleftrightarrow\lambda\lambda$ is shown to be nontrivial.

Following the discussion in Section~\ref{subsec:Junctions} we also exploited the robustness of the single junction structure to experimentally construct a topologically stable network with $(\chi,\chi,-1)$ defect junctions as building blocks (Fig.~\ref{fig:network}). To create the defect network, we defined surface boundary conditions that lead to an array of $\pi$-twist walls. Specifically, larger distances between $\pi$-twist walls prefer forming $\chi$ defects while a $2\pi$-twist wall generates a $-1$ line instead. By controlling the width and separation distances of the twist walls, manipulation of the surface pattern controls the formation of defect junctions (Methods). 
As presented in Fig.~\ref{fig:network}, the $\chi$ and $-1$ defects are characterized and easily distinguished microscopically.
The experimental realization of the network is expandable by including more repeating units of the surface pattern, reconfigurable by employing laser tweezers, and most importantly, always topologically stable as guaranteed by the non-Abelian biaxial topology of the fundamental group elements of CLCs.

\section{Energetic stability of a bound state of two defect lines}\label{sec:boundstate}

Building on the approach in Ref.~\cite{monastyrskii_bound_2006}, we now analyze the bound state of two parallel defect lines. This is done within the elastic model given by Eq.~(\ref{Qenergy}).
Bound states of this kind are seen experimentally and numerically in junctions such as Fig.~\ref{fig:junction_sim}(a) and Fig.~\ref{fig:junction_exp}. 

Suppose we have two defect lines with topological charge $q_1$ and $q_2$ separated by  distance $\rho$ in a cylindrical region $\Omega = D_R\times [0, L]$, where $D_R$ is a disc of radius $R$. The region of interest away from the defect core with radius $r_c$ is the annular region $r_c\ll r\ll R$. Here, the $Q$-tensor may be continuously deformed to satisfy the sliding boundary condition $\partial_rQ_{ij}=0$, which allows us to assume that the order parameter chiefly depends on the azimuthal angle. The elastic free energy density can then be approximated by
    \begin{equation}
        f_{elastic} \approx \frac{f(\phi)}{r^2}+\frac{g(\phi)}{r} \ ,
    \end{equation}
where $f(\phi)$ is the angular dependence of the expansion of the $\gamma_1$,$\gamma_2$, and $\gamma_6$ terms in Eq.~(\ref{Qenergy}). Similarly, $g(\phi)$ is the angular dependence of the chiral $\gamma_4$ term in Eq.~(\ref{ChiralTerm}). The free energy can now be approximated by integrating over the volume of $\Omega$ which, for convenience, can be split into three terms:  
    \begin{equation}
        \mathcal{F} = \mathcal{F}_{q_1} + \mathcal{F}_{q_2} + \mathcal{F}_{q_3} \ .
    \end{equation}
The first two terms are identical up to the interchange $q_1\leftrightarrow q_2$ and represent the individual energies of each defect. The radial integrals for these two terms are calculated in the annulus $r_c<r<\rho$ surrounding each defect line and take the form
    \begin{align}
        \mathcal{F}_{q_{1,2}} &\approx \int_{0}^Ldz\,\int_{0}^{2\pi}d\phi\,\int_{r_c}^\rho rdr\,\left(\frac{f(\phi)}{r^2}+\frac{g(\phi)}{r}\right) \\
        &= \gamma_1 L S^2 K_{q_{1,2}}^{(1)}\ln\left(\frac{\rho}{r_c}\right) + \gamma_4 LS^2 (\rho-r_c) K_{q_{1,2}}^{(6)} \ .
     \end{align}
Similarly, the third term is computed over the annular region $\rho<r<R$ surrounding the bound state 
    \begin{align}
        \mathcal{F}_{q_{3}} &\approx \int_{0}^Ldz\,\int_{0}^{2\pi}d\phi\,\int_{\rho}^Rrdr\,\left(\frac{f(\phi)}{r^2}+\frac{g(\phi)}{r}\right) \\
        &= \gamma_1 L S^2 K_{q_{3}}^{(1)}\ln\left(\frac{R}{\rho}\right) + \gamma_4 LS^2 (R-\rho) K_{q_{3}}^{(4)},
    \end{align}
where we have introduced the elastic constants associated with each defect class. They are obtained from the azimuthal integrals:
    \begin{align}
        \bar{K}_{q_i}^{(1)} &\equiv F_{q_i}^{(1)}\left(\frac{T}{S},\frac{\gamma_2}{\gamma_1}\right) + \frac{\gamma_6}{\gamma_1} S F_{q_i}^{(6)}\left(\frac{T}{S}\right) \\
        &\equiv K_{q_i}^{(1)}+\frac{\gamma_6}{\gamma_1} S K_{q_i}^{(6)} \\
        K_{q_i}^{(4)} &\equiv G_{q_i}\left(\frac{T}{S}\right)\label{K4} ,
    \end{align}
     
where the functions $F_{q_i}^{(1)}$ and $F_{q_i}^{(6)}$ are the azimuthal contributions of the $\gamma_1,\gamma_2$ terms and $\gamma_6$ term, respectively. Similarly, $G_{q_i}$ is the azimuthal contribution of the $\gamma_4$ term. Additionally, upon defining the elastic energy losses:
    \begin{align}
        \Delta \bar{K}^{(1)} &\equiv \bar{K}_{q_1}^{(1)}+\bar{K}_{q_2}^{(1)}-\bar{K}_{q_3}^{(1)} \\
        \Delta K^{(4)} &\equiv K_{q_1}^{(4)}+K_{q_2}^{(4)}-K_{q_3}^{(4)}.
    \end{align}
The energy then becomes:
    \begin{align}
        \mathcal{F} &= \gamma_1 S^2L\bigg[\Delta\bar{K}^{(1)}\ln\left(\frac{\rho}{r_c}\right) + K_{q_3}^{(1)}\ln\left(\frac{R}{r_c}\right)\bigg]  \nonumber\\
         &\hspace{10pt}+\gamma_4S^2L\bigg[\Delta K^{(4)}(\rho-r_c) + K_{q_3}^{(4)}(R-r_c)\bigg].
         \label{chiralEnergy}
    \end{align}
The minimum energy occurs at the separation
    \begin{equation}
        \rho_{min} = -\frac{\gamma_1\Delta\bar{K}^{(1)}}{\gamma_4\Delta K^{(4)}}
    \end{equation}
and the second derivative at this point is 
    \begin{align}
        \pdv[2]{\mathcal{F}}{\rho}\bigg|_{\rho_{min}} = -\frac{\gamma_4^2(\Delta K^{(4)})^2}{\gamma_1\Delta \bar{K}^{(1)}} \ ,
    \end{align}
which requires $\Delta\bar{K}^{(1)}<0$, corresponding to defect repulsion, for $\rho_{min}$ to be a minimum.  

Recall that the sign of $\gamma_4$ determines the handedness of the system. For $\gamma_4>0$, $\rho_{min}>0$ requires $\Delta K^{(4)}>0$. On the other hand, if $\gamma_4<0$, we must have $\Delta K^{(4)}<0$. Seemingly, the sign of the $\Delta K^{(4)}$ term depends on the handedness. If we instead focus on the sign of the product $\gamma_4\Delta K^{(4)}$, we see that it must be positive in both cases. This must be the case because unlike the constants $\bar{K}_{q_i}^{(1)}$, which have some memory of the elastic moduli $\gamma_1$ and $\gamma_6$, the $K_{q_i}^{(4)}$ do not, as seen in Eq.~(\ref{K4}). Thus, it is more appropriate to consider the sign of $\gamma_4\Delta K^{(4)}>0$ for a minimum to exist in the repulsive case, with $\Delta\bar{K}^{(1)}<0$. 

\begin{figure}
        \centering 
        \includegraphics[width=\columnwidth]{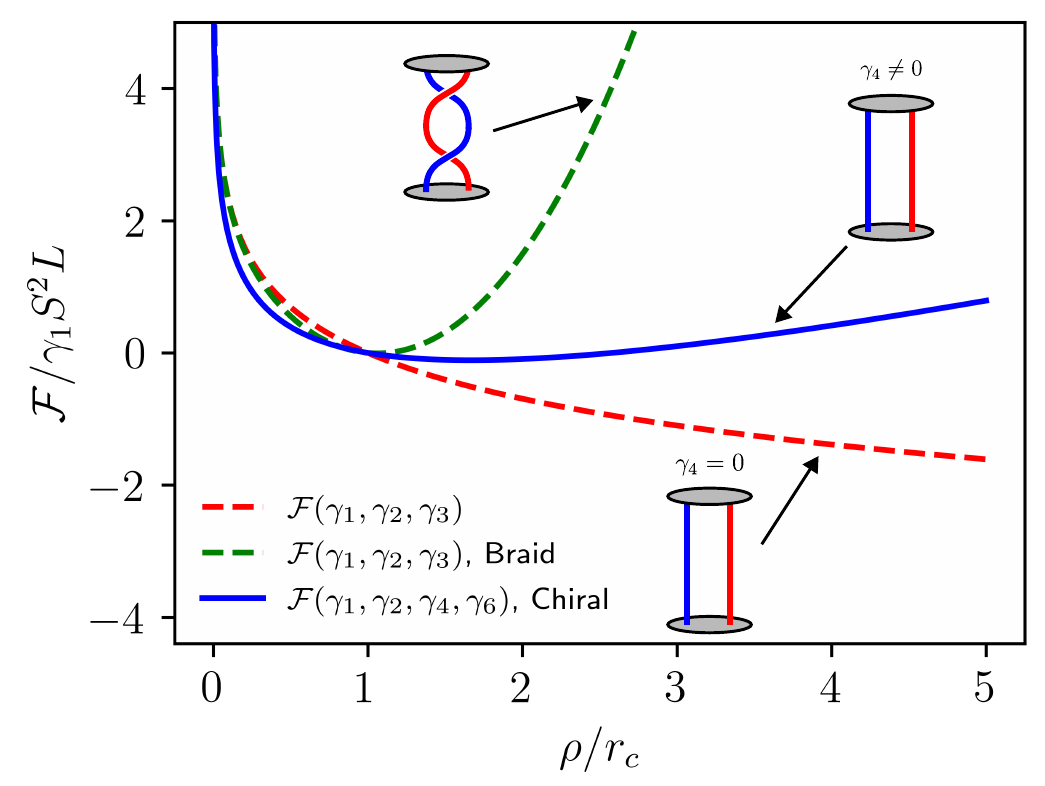}
        \caption{Elastic free energy for three different bound state configurations: (1) achiral, unbraided (red, dashed); (2) chiral, unbraided (blue, solid); (3) achiral, braided (green,  dashed). Of the three, only the achiral, unbraided case does not have a stable minimum radius.
        }
        \label{fig:EnergyPlot}
\end{figure}

One can check that upon writing the elastic constants $K_{q_i}^{(1)}$ in terms of the Frank elastic constants in Eq.~(\ref{FrankOseenE}) and using the mapping described in Appendix \ref{apdxB} to relate to the biaxial coupling constants, $\gamma_i$, the repulsive condition is satisfied subject to labeling the defects corresponding to $C_\lambda$ with the smallest $K_{q_i}^{(1)}$. Doing the same for $K_{q_i}^{(4)}$ reveals that this condition is met when $T/S<2/(\sqrt{5}-1)$.
 
Unlike the achiral non-anisotropic case, where the repulsive condition leads to the energy being minimized when the defects are as far apart as possible, here we have an energy-minimizing separation distance that is stabilized within the bounds of the medium by none other than the elastic tension provided by the intrinsic chiral nature of the bulk. Fig.~\ref{fig:EnergyPlot} shows the energy in Eq.~\ref{chiralEnergy} along with the achiral braided and unbraided cases. This agrees with the observation of bound states of $\lambda$ and $\tau$ defect pairs in the D2CLCs. 

We can further compare this observation to the braided achiral non-anisotropic case in which there is a stable minimum radius. The tension that creates the bound state in that case is provided by an extrinsic chirality generated by physically braiding the two defects.

\section{Conclusions}

The non-Abelian properties of the three-dimensional line defects have been of interest for decades but rarely seen in experiments 
\cite{hall2016tying,annala2022topologically,rajamaki2023topologically, kobayashi_physics_2011}.  Here we uncover new metastructural features of such systems and realize them in the laboratory in a chiral liquid crystal with orthorhombic biaxial symmetry breaking.    

In particular we show that the four non-trivial classes of line defects have several distinct realizations: they may be standalone elementary defects, bound states of two elementary defects as well as elements of junctions, both single and extended networks, all consistent with the well-known quaternionic algebra that governs their classification and possible defect couplings. 

In this way we reveal features of such systems beyond those obvious from their well-known mathematical classification and 
point the way to the construction of even more elaborate topologically-rigid structures that pose theoretical and  experimental challenges and may well have technological utility. The demonstrated experimental embodiments of defect networks not only comply with topological rules and constrains but also fearture low-free-energy states, showing how our specific physical system selects lowest-energy configurations out of all allowed by topology rules.  
In addition to the rich topological behaviors of D2CLCs realized in our specific material system, future research on materials with atypical elasticities, such as CB7CB \cite{tai2022geometric}, could reveal other energetic pathways of nonabelian algebra as well as defect junctions and networks that haven't yet been observed experimentally.



\section{Acknowledgements}
The experimental and numerical simulations research at University of Colorado Boulder was supported by the US Department of Energy, Office of Basic Energy Sciences, Division of Materials Sciences and Engineering, under contract DE-SC0019293 with the University of Colorado at Boulder. I.I.S. acknowledges the support of the International Institute for Sustainability with Knotted Chiral Meta Matter (WPI-SKCM2) at Hiroshima University in Japan during part of his sabbatical stay when this article was under preparation. I.I.S. and J.-S.W. acknowledge the hospitality of the Kavli Institute for Theoretical Physics in Santa Barbara with support by the National Science Foundation under Grant No. NSF PHY-2309135. R.A.V. acknowledges support in part by the National Science Foundation under Grant No. NSF PHY-2309135 as well as the National Science Foundation California LSAMP Bridge to the Doctorate Fellowship under Grant No. HRD-1701365.

\appendix
\section{Fundamental group of systems with biaxial symmetry}

The fact that biaxial nematic systems have disclinations whose algebra is that of the quaternion group, $Q_8$, is well known and quoted with confidence. However, for the sake of self-containment, here we prove the statement. In a more mathematical language, we want to prove that
\begin{equation}
    \pi_1(SO(3)/D_2) \simeq Q_8.
    \label{pi1_Q}
\end{equation}
To do so, we require the following theorem which proves useful in computing homotopy groups of coset spaces:
\begin{theorem}
    Let $G$ be a simply connected Lie group with subgroup $H\leq G$ and identity component $H_0\trianglelefteq G$. Then 
    \begin{equation}
        \pi_1(G/H)\simeq H/H_0
    \end{equation}
    \label{thm:theorem1}
\end{theorem}

Proof of this statement establishes an isomorphism between the fundamental group and the quotient space $H/H_0$ by relating the loops in $G/H$ based at $H$ to paths in $G$ from a connected piece of $H$ to the piece that contains the identity, $H_0$.

Theorem~\ref{thm:theorem1} allows one to compute $\pi_1(SO(3)/D_2)$ by lifting to a universal cover map. This is 
    \begin{align}
        SO(3) &\longrightarrow SU(2) \\
        D_2 &\longrightarrow Q_8
        \label{covering map}
    \end{align}
Here, the covering map sends $\pi$ rotations about each symmetry axis of the rectangle, into $\pi$ rotations in $SU(2)$ parametrized by the Pauli matrices. The set of such rotations forms the lift of $D_2$, that is
    \begin{equation}
        \{\pm1,\pm i\sigma_x,\pm i\sigma_y,\pm i\sigma_z\}
        \label{SU2set}
    \end{equation}
which is simply the group of quaternions, $Q_8$.

Now, applying theorem 1, we have $\pi_1(SO(3)/D_2)\simeq\pi_1(SU(2)/Q_8)\simeq Q_8/(Q_8)_0 = Q_8$ since $(Q_8)_0 = \{1\}$.

\section{Relation between the elastic moduli of the biaxial Q-tensor model and those of the vectorial model}
\label{apdxB}
Following the approach similar to Ref.~\cite{mori1999multidimensional}, the elastic constants in the Q-tensor model and the vectorial model, which describes energy in terms of gradients of director fields, can be related by expanding Eq.~\ref{Qenergy} and Eq.~\ref{FrankOseenE} and collecting terms with the same invariants, which serve as linear-independent bases for the projection. Ignoring the surface terms (those with the form $\nabla \cdot f$~\cite{govers_elastic_1984}), the equations read:

    \begin{align}
        K_1 &= \frac{2}{3} S \left[ (S-T)(6\gamma_1+3\gamma_2\hide{+3\gamma_3}+2T\gamma_6) - 2S^2\gamma_6 \right] \nonumber \\
        K_2 &= \frac{4}{3} S \left[ (S-T)(3\gamma_1+T\gamma_6) - S^2\gamma_6 \right] \nonumber \\
        K_3 &= \frac{2}{3} S \left[ (S-T)(6\gamma_1+3\gamma_2\hide{+3\gamma_3}+2T\gamma_6) - 6ST\gamma_6 \right] \nonumber \\
        K_4 &= -\frac{2}{3} T \left[ (S-T)(6\gamma_1+3\gamma_2\hide{+3\gamma_3}-2T\gamma_6) + 2S^2\gamma_6 \right] \nonumber \\
        K_5 &= -\frac{4}{3} T \left[ 3(S-T)\gamma_1 + (S^2-ST+T^2)\gamma_6\right] \nonumber \\
        K_6 &= -\frac{2}{3} T \left[ (S-T)(6\gamma_1+3\gamma_2\hide{+3\gamma_3}) + 2S(S+2T)\gamma_6 \right] \nonumber \\
        K_7 &= 4ST(S-T)\gamma_6 \nonumber \\
        K_8 &= 4ST(T-S)\gamma_6 \nonumber \\
        K_9 &= 2ST(\gamma_2\hide{+\gamma_3}-2S\gamma_6) \nonumber \\
        K_{(10)} &= 2ST(\gamma_2\hide{+\gamma_3}-2T\gamma_6) \nonumber \\
        K_{(11)} &= \frac{4}{3}ST \left[ 3\gamma_1+(S+T)\gamma_6 \right] \nonumber \\
        K_{(12)} &= \frac{2}{3}ST \left[ 6\gamma_1+3\gamma_2\hide{+3\gamma_3}+2(S+T)\gamma_6 \right] 
    \end{align}
Similarly, the chiral twisting parts~\ref{FOchiral} have:
    \begin{align}
        K_{(13)} &= S^2\gamma_4 \nonumber \\
        K_{(14)} &= T^2\gamma_4 \nonumber \\
        K_{(15)} &= K_{(16)} = K_{(17)} = ST\gamma_4 
    \end{align}

Since the intrinsic biaxiality of our chiral liquid crystal is much smaller than its uniaxiality order parameter $T \ll S$, we can further calculate the coefficients $\gamma_i, i=1,2,6$ as:
    \begin{align}
        \gamma_1 &= \frac{1}{4S^2}(K_2+K_3-K_1) \nonumber \\
        \gamma_2\hide{+\gamma_3} &= \frac{1}{2S^2}(K_1-K_2) \nonumber \\
        \gamma_6 &= \frac{3}{4S^3}(K_3-K_1) \nonumber \\
    \end{align}
which give an estimation of the orthorhombic biaxial elastic moduli $K_i,i=4-12$ up to $\mathcal{O}(T/S)$ for a CLC:
    \begin{align}
        K_4 &= (K_1-2K_3)\frac{T}{S} \nonumber \\
        K_5 &= (K_2+2K_3-2K_1)\frac{T}{S} \nonumber \\
        K_6 &= (K_1-2K_3)\frac{T}{S} \nonumber \\
        K_7 &= 3(K_3-K_1)\frac{T}{S} \nonumber \\
        K_8 &= 3(K_1-K_3)\frac{T}{S} \nonumber \\
        K_9 &= (K_2+3K_3-4K_1)\frac{T}{S} \nonumber \\
        K_{(10)} &= (K_1-K_2)\frac{T}{S} \nonumber \\
        K_{(11)} &= (K_2+2K_3-2K_1)\frac{T}{S} \nonumber \\
        K_{(12)} &= (2K_3-K_1)\frac{T}{S}
    \end{align}
where $K_1,K_2,K_3$ are the elastic constants for splay, twist, and bend deformations respectively. Given the limited types of allowed deformations as in Eq.~\ref{Qenergy}, we can estimate the biaxial elasticities based on the uniaxial moduli and scalar order parameters of chiral nematic systems.

Taking a 5CB-dominated LC with a cholesteric pitch $p=5\um$, we estimated the 12 elastic moduli and compared them to another studied biaxial LC system~\cite{senyuk2021nematoelasticity}:

\begin{table}[!h]
    \centering
    \begin{tabular}{c|c|c}
        elastic constants (pN) & chiral nematic LC & hybrid LC \cite{senyuk2021nematoelasticity} \\
        \hline \hline
        $K_{1}$ & 6 & 6.15\\
        $K_{2}$ & 3 & 3\\
        $K_{3}$ & 10 & 10\\
        $K_{4}$ & $\sim0$ & $\sim0$\\
        $K_{5}$ & $\sim0$ & $\sim0$\\
        $K_{6}$ & $\sim0$ & 0.3\\
        $K_{7}$ & 0.02 & 6.45\\
        $K_{8}$ & $\sim0$ & 10\\
        $K_{9}$ & $\sim0$ & 6.15\\
        $K_{(10)}$ & 0.005 & $\sim0$\\
        $K_{(11)}$ & 0.018 & 3\\
        $K_{(12)}$ & 0.023 & 10.3
    \end{tabular}
    \caption{Biaxial elastic moduli estimation for two biaxial systems both derived from nematic 5CB, one doped with a chiral additive and the other with anisotropic  nanoparticles to form a molecular-colloidal hybrid LC.}
    \label{tab:elasticities_values}
\end{table}

As one would expect, weaker biaxiality compared to hybrid systems implies weaker biaxial elasticities. Manufacturing technologies associated with CLCs are, however, much more developed for exploring the biaxial topologies in soft matter systems.

\clearpage

\bibliographystyle{apsrev4-1}
\bibliography{refs}

\end{document}